\documentclass[3p]{elsarticle}

\usepackage{amsmath}
\usepackage{amsthm}
\usepackage{amssymb}
\usepackage{graphicx}
\usepackage{esint}
\usepackage{color}

\usepackage{lineno}
\PassOptionsToPackage{hyphens}{url}
\usepackage[pdfencoding=auto,psdextra]{hyperref}
\usepackage{bookmark}

\biboptions{numbers,sort&compress}

\modulolinenumbers[1]

\journal{Chaos, Solitons \& Fractals}

\begin{document}

\begin{frontmatter}

\title{Quantum tunneling of Davydov solitons through massive barriers}

\author[address1]{Danko D. Georgiev\corref{mycorrespondingauthor}}
\ead{danko.georgiev@mail.bg}
\cortext[mycorrespondingauthor]{Corresponding author}

\author[address2]{James F. Glazebrook}
\ead{jfglazebrook@eiu.edu}

\address[address1]{Institute for Advanced Study, Varna, Bulgaria}
\address[address2]{Department of Mathematics and Computer Science, Eastern Illinois University, Charleston, IL 61920, USA}

\begin{abstract}
Protein clamps provide the cell with effective mechanisms for sensing of environmental changes and triggering adaptations that maintain homeostasis. The general physical mechanism behind protein clamping action, however, is poorly understood. Here, we explore the Davydov model for quantum transport of amide~I energy, which is self-trapped in soliton states that propagate along the protein $\alpha$-helix spines and preserve their shape upon reflection from the $\alpha$-helix ends. We study computationally whether the Davydov solitons are able to reflect from massive barriers that model the presence of external protein clamps acting on a portion of the $\alpha$-helix, and characterize the range of barrier conditions for which the Davydov solitons are capable of tunneling through the barrier. The simulations showed that the variability of amino acid masses in proteins has a negligible effect on Davydov soliton dynamics, but a massive barrier that is a hundred times the mass of a single amino acid presents an obstacle for the soliton propagation. The greater width of Davydov solitons stabilizes the soliton upon reflection from such a massive barrier and increases the probability of tunneling through it, whereas the greater isotropy of the exciton-lattice coupling suppresses the tunneling efficiency by increasing the interaction time between the Davydov soliton and the barrier, thereby prolonging the tunneling time through the barrier and enhancing the probability for reflection from the latter. The results as presented, demonstrate the feasibility of Davydov solitons reflecting from or tunneling through massive barriers, and suggest a general physical mechanism underlying the action of protein clamps through local enhancement of an effective protein $\alpha$-helix mass.
\end{abstract}


\date{April 9, 2019}

\end{frontmatter}

%\linenumbers

\section{Introduction}

Proteins are the molecular nano-engines that support all activities essential to life \cite{Strong2004,Hess2005,Hirokawa2010,Ha2016}. Protein-protein interactions regulate key biochemical processes, including DNA replication, repair and transcription, mRNA translation, protein sorting and transport, metabolic pathways, cellular signaling, sensing of environmental changes and triggering adaptations that maintain homeostasis \cite{Jones1996,Ottman2013,Keskin2016}. Functional protein complexes are often clamped in an inactive meta-stable state that can be readily activated by signaling cascades triggering the release of the inhibitory clamp. Neurotransmission at active synapses in the brain is an example of a biological process that is tightly regulated by protein clamps, such as synaptotagmin and complexin \cite{Giraudo2006}, which inhibit the SNARE protein complex that drives synaptic vesicle exocytosis in the absence of electric spikes, but de-inhibit and assist the SNARE complex function upon electric excitation of the presynaptic axonal button \cite{GeorgievGlazebrook2007,GeorgievGlazebrook2012,Georgiev2017}.
Protein clamps that inhibit electric excitability and the flow of Ca$^{2+}$ currents through neuronal membrane-bound receptors \cite{Li2013} or voltage-gated ion channels \cite{Simms2014} are also known.
Predicting protein--protein interactions \cite{Hashemifar2018} and elucidating the mechanisms behind protein clamping action \cite{Douma2017,Iyer2013}, are important open problems currently under investigation.

In the quest for a general physical mechanism behind protein clamping action, the aim of this present work is to study the transport of amide I vibrational energy in proteins as propagated by Davydov solitons \cite{Davydov1976,Davydov1979,Davydov1982,Davydov1986,Davydov1988,Scott1984,Scott1985,Scott1992,Brizhik2004,Brizhik2006,Brizhik2010}, where the presence of massive barriers is taken into account. For a single $\alpha$-helix spine of
hydrogen-bonded peptide groups, the generalized Davydov Hamiltonian is
\begin{gather}
\hat{H}=\sum_{n} \left[E_{0}\hat{a}_{n}^{\dagger}\hat{a}_{n}-J\left(\hat{a}_{n}^{\dagger}\hat{a}_{n+1}+\hat{a}_{n}^{\dagger}\hat{a}_{n-1}\right)\right]
+\frac{1}{2} \sum_{n} \left[\frac{\hat{p}_{n}^{2}}{M_{n}}+w\left(\hat{u}_{n+1}-\hat{u}_{n}\right)^{2}\right]\nonumber\\
+\chi \sum_{n}\left[\hat{u}_{n+1}+\left(\xi-1\right)\hat{u}_{n}-\xi\hat{u}_{n-1}\right]\hat{a}_{n}^{\dagger}\hat{a}_{n}
\label{eq:H}
\end{gather}
where the index $n$ counts the peptide groups along the $\alpha$-helix spine, $\hat{a}_{n}^{\dagger}$ and $\hat{a}_{n}$ are the boson creation and annihilation operators respectively for the amide~I exciton, $E_{0}=32.8$~zJ (zeptojoule) is the amide~I exciton energy, $J=0.155$~zJ is the nearest neighbor exciton dipole-dipole
coupling energy along the spine, $M_{n}$ is the mass, $\hat{p}_{n}$ is the momentum operator and $\hat{u}_{n}$ is the displacement operator from the equilibrium position of the $n$th peptide group, $w$ is the spring constant of the hydrogen bonds in the lattice \cite{Davydov1976,Davydov1979,Davydov1982,Davydov1986,Davydov1988,Scott1984,Scott1985,Scott1992}, $\chi=\bar{\chi}\frac{2}{1+\xi}$ is an anharmonic parameter arising from the coupling between the amide~I exciton and the phonon lattice displacements, $\bar{\chi}=\frac{\chi_{r}+\chi_{l}}{2}$ is the average of the right $\chi_{r}$ and left $\chi_{l}$ coupling parameters, and $\xi=\frac{\chi_{l}}{\chi_{r}}$ is the isotropy parameter of the exciton-lattice coupling ($\chi_{r}\neq0$ and $0\leq\chi_{l}\leq\chi_{r}$, hence $\xi$ varies in the interval $\left[0,1\right]$) \cite{Luo2017,GeorgievGlazebrook2019}.

The quantum equations of motion can be derived from the Hamiltonian \eqref{eq:H} with the use of the second of Davydov's ansatz state vectors \cite{Davydov1982,Sun2010,Luo2011,Zhou2015}
\begin{equation}
|D_{2}(t)\rangle=\sum_{n}a_{n}(t)\hat{a}_{n}^{\dagger}|0_{\textrm{ex}}\rangle e^{-\frac{\imath}{\hbar}\sum_{j}\left(b_{j}(t)\hat{p}_{j}-c_{j}(t)\hat{u}_{j}\right)}|0_{\textrm{ph}}\rangle
\end{equation}
which provides an excellent approximation \cite{Sun2010,GeorgievGlazebrook2019} to the exact solution of the Schr\"{o}dinger equation
\begin{equation}
\imath\hbar\frac{d}{dt}|D_{2}(t)\rangle=\hat{H}|D_{2}(t)\rangle\label{eq:s}
\end{equation}
The expectation value of the exciton number operator $\hat{N}_n=\hat{a}_{n}^{\dagger}\hat{a}_{n}$ at the peptide group $n$ is
\begin{equation}
\langle\hat{N}_n\rangle = \langle D_{2}(t)|\hat{N}_n|D_{2}(t)\rangle=|a_{n}|^{2}
\end{equation}
The total probability of finding the amide~I excition inside the protein should be normalized, $\sum_n |a_n|^2=1$. The latter condition ensures that the Davydov's ansatz is normalized as well, $\langle D_{2}(t)|D_{2}(t)\rangle=1$.

With the use of the Hadamard lemma
\begin{equation}
e^{\hat{A}}\hat{B}e^{-\hat{A}}
=\hat{B}+\left[\hat{A},\hat{B}\right]+\frac{1}{2!}\left[\hat{A},\left[\hat{A},\hat{B}\right]\right]+\frac{1}{3!}\left[\hat{A},\left[\hat{A},\left[\hat{A},\hat{B}\right]\right]\right]+\ldots
\end{equation}
the expectation values of the phonon displacement and momentum operators, $\hat{u}_{n}$ and
$\hat{p}_{n}$, are found to be
\begin{eqnarray}
\langle\hat{u}_{n}\rangle & = & \langle D_{2}(t)|\hat{u}_{n}|D_{2}(t)\rangle=b_{n}\\
\langle\hat{p}_{n}\rangle & = & \langle D_{2}(t)|\hat{p}_{n}|D_{2}(t)\rangle=c_{n}
\end{eqnarray}
Applying the generalized Ehrenfest theorem for the time dynamics
of the expectation values
\begin{eqnarray}
\imath\hbar \frac{d}{dt}b_{n} & = & \langle\left[\hat{u}_{n},\hat{H}\right]\rangle\\
\imath\hbar \frac{d}{dt}c_{n} & = & \langle\left[\hat{p}_{n},\hat{H}\right]\rangle
\end{eqnarray}
together with the Schr\"{o}dinger equation \eqref{eq:s}, leads
to the following system of gauge transformed quantum equations of
motion (for detailed derivation see \cite{GeorgievGlazebrook2019,Kerr1987, Kerr1990}):
\begin{eqnarray}
\imath\hbar\frac{da_{n}}{dt} & = & -J\left(a_{n+1}+a_{n-1}\right) + \chi\left[b_{n+1}+(\xi-1)b_{n}-\xi b_{n-1}\right]a_{n}\\
M_{n}\frac{d^{2}}{dt^{2}}b_{n} & = & w\Big(b_{n-1}-2b_{n}+b_{n+1})-\chi\Big(\left|a_{n-1}\right|^{2}+(\xi-1)\left|a_{n}\right|^{2}-\xi\left|a_{n+1}\right|^{2}\Big)
\end{eqnarray}
where $a_{n}$ is the quantum probability amplitude of the amide~I exciton at the $n$th site and $b_{n}$ is the expectation value of the longitudinal displacement from the equilibrium position of the $n$th peptide group.

\section{Computational study}

\subsection{Model parameters}

We have computationally simulated the system of Davydov equations for a protein $\alpha$-helix spine with $n=40$ peptide groups, $w=13$~N/m \cite{Itoh1972,MacNeil1984,Scott1984,Cruzeiro1988} and $\bar{\chi}=35$ pN (piconewton) \cite{Scott1992}.
The method for numerical integration was LSODA, Livermore Solver for Ordinary Differential equations with Automatic selection between nonstiff (Adams) and stiff (Backward Differentiation Formula, BDF) methods, developed by Hindmarsh and Petzold \cite{Petzold1983,Hindmarsh1983,Hindmarsh1995,Trott2006}. LSODA implements an algorithm that uses the nonstiff method initially, then followed by dynamical monitoring of the data in order to decide which method to use at the end of each step of the integration. If the problem changes character (i.e., from nonstiff to stiff or vice versa) in the interval of integration, the solver automatically switches to the method (Adams or BDF) which is likely to be most efficient for that part of the problem \cite{Petzold1983}.
Solitons with different widths were produced by different normalized initial Gaussian distributions $\sigma$ of amide~I energy spread over 1, 3, 5 or 7 peptide groups such that the corresponding quantum probability amplitudes for the non-zero $a_{n}(0)$ were: $\{1\}$, $\{\sqrt{0.244},\sqrt{0.512},\sqrt{0.244}\}$, $\{\sqrt{0.099},\sqrt{0.24},\sqrt{0.322},\sqrt{0.24},\sqrt{0.099}\}$ and $\{\sqrt{0.059},\sqrt{0.126},\sqrt{0.199},\sqrt{0.232},\sqrt{0.199},\sqrt{0.126},\sqrt{0.059}\}$. The lattice of hydrogen bonds was initially unperturbed, $b_n(0)=0$ and $\frac{d b_n(0)}{dt}=0$ \cite{GeorgievGlazebrook2019}. To model a real protein $\alpha$-helix, the peptide groups at the N-end or C-end were coupled to a single neighbor, that is, there were no peptide groups with index $n=0$ or $n=41$. This effectively creates reflective boundaries at the protein $\alpha$-helix ends. It had been previously shown that the same set of initial conditions for the amide~I exciton and phonon lattice leads to stationary solitons when the protein $\alpha$-helix is modeled with periodic boundary conditions that effectively compactify the $\alpha$-helix spine into a circle \cite{GeorgievGlazebrook2019}.

Previous studies of Davydov solitons modeled the amino acid residues
as having equal mass, taken to be the average mass of an amino acid
inside the protein $\alpha$-helix, namely, $M=0.19$ zg (zeptogram).
This modeling assumption is feasible because the biochemical variability
of amino acid masses is in the range $1\pm0.64M$ (Table \ref{tab:1})
and has minor effects on the dynamics of the soliton \cite{Motschmann1989,Forner1990,Forner1991c}.

\begin{figure}
\begin{centering}
\includegraphics[width=160mm]{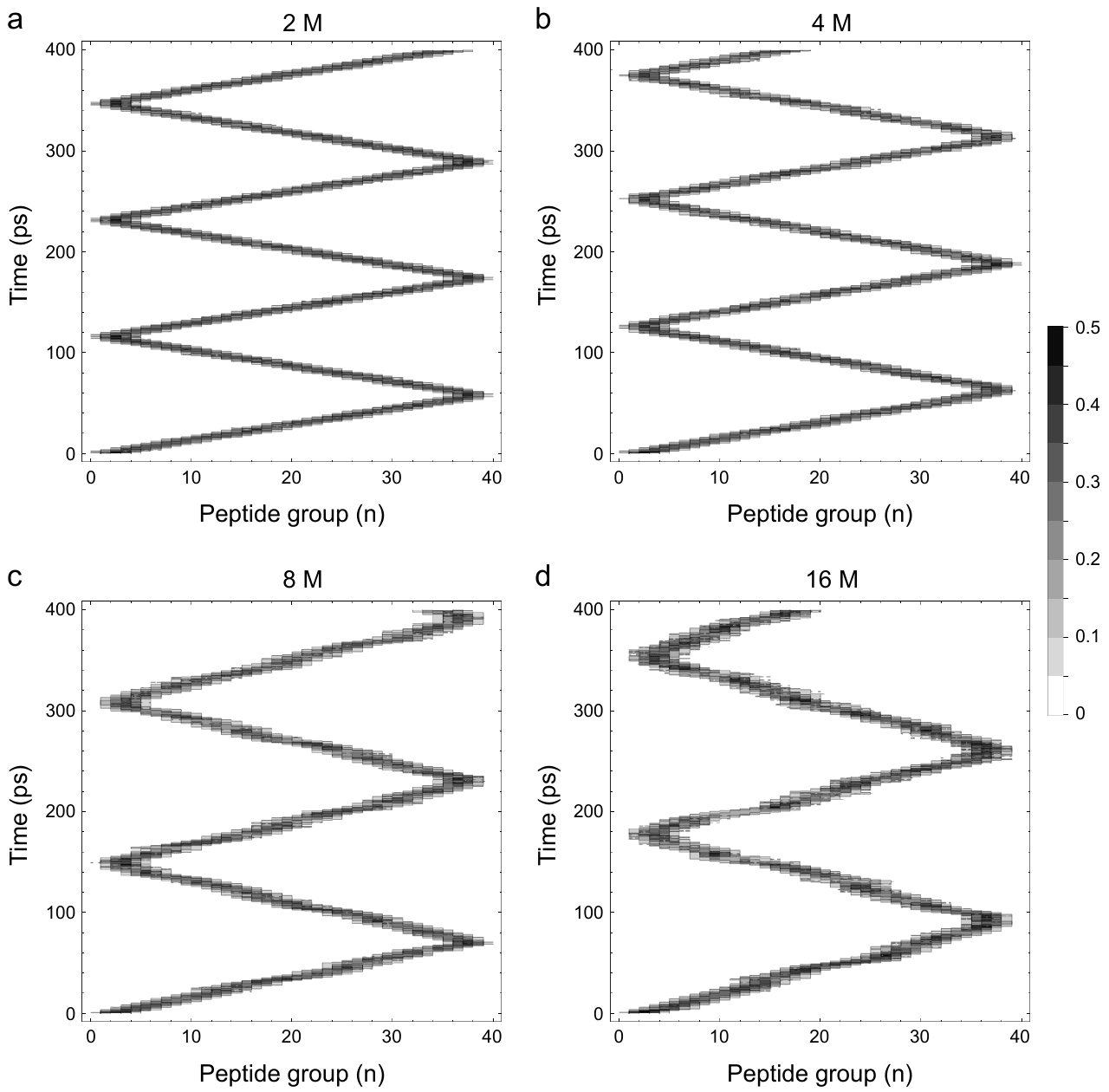}
\par\end{centering}
\caption{\label{fig:1}Soliton dynamics visualized through $|a_{n}|^{2}$ for
different masses of the amino acid residues $2M$, $4M$, $8M$ or
$16M$, for the isotropic exciton-lattice coupling $\xi=1$ launched by a Gaussian of amide~I energy
$\sigma=5$ applied at the N-end of an $\alpha$-helix spine composed
of 40 peptide groups during a period of 400 ps.}
\end{figure}

Our simulations, in which the mass of all amino acid residues was doubled
iteratively, showed that increasing the amino acid mass outside of
the biochemical variability range decreases the average soliton speed~$v$.
As an example, a Davydov soliton with $\xi=1$ whose width is
$\sigma=5$ propagates with speed $v=314$ m/s for $M_{n}=1M$, $v=298$
m/s for $M_{n}=2M$, $v=269$ m/s for $M_{n}=4M$, $v=235$ m/s for
$M_{n}=8M$, and $v=178$ m/s for $M_{n}=16M$ (Fig.~\ref{fig:1}).
The increased amino acid mass also introduces fluctuations of the
soliton speed at a timescale of $10-20$ ps (Fig.~\ref{fig:1}d) due
to stronger interaction with the lattice phonon waves. Taking into consideration these
preliminary simulations, we have concluded that setting the amino acid
residues as having equal mass of $1M$ provides a good base model
on top of which can be added massive barriers towards representing external
protein clamping action on the $\alpha$-helix.

\begin{table}
\caption{\label{tab:1}Properties of amino acid residues in protein $\alpha$-helices.}

\begin{centering}
\begin{tabular}{|c|c|c|c|c|}
\hline
Amino acid & Code & Helical penalty (kJ/mol) \cite{Pace1998} & Mass (Da) & Mass ($M=0.19$ zg)\tabularnewline
\hline
\hline
Alanine & A & 0.00 & 71 & 0.63\tabularnewline
\hline
Arginine & R & 0.88 & 156 & 1.37\tabularnewline
\hline
Leucine & L & 0.88 & 113 & 0.99\tabularnewline
\hline
Methionine & M & 1.00 & 131 & 1.15\tabularnewline
\hline
Lysine & K & 1.09 & 128 & 1.13\tabularnewline
\hline
Glutamine & Q & 1.63 & 128 & 1.13\tabularnewline
\hline
Glutamate & E & 1.67 & 129 & 1.14\tabularnewline
\hline
Isoleucine & I & 1.72 & 113 & 0.99\tabularnewline
\hline
Tryptophan & W & 2.05 & 186 & 1.64\tabularnewline
\hline
Serine & S & 2.09 & 87 & 0.77\tabularnewline
\hline
Tyrosine & Y & 2.22 & 163 & 1.43\tabularnewline
\hline
Phenylalanine & F & 2.26 & 147 & 1.29\tabularnewline
\hline
Histidine & H & 2.55 & 137 & 1.21\tabularnewline
\hline
Valine & V & 2.55 & 99 & 0.87\tabularnewline
\hline
Asparagine & N & 2.72 & 114 & 1.00\tabularnewline
\hline
Threonine & T & 2.76 & 101 & 0.89\tabularnewline
\hline
Cysteine & C & 2.85 & 103 & 0.91\tabularnewline
\hline
Aspartate & D & 2.89 & 115 & 1.01\tabularnewline
\hline
Glycine & G & 4.18 & 57 & 0.50\tabularnewline
\hline
Proline & P & 13.22 & 97 & 0.85\tabularnewline
\hline
\end{tabular}
\par\end{centering}

\end{table}

\subsection{Effect of soliton width on reflection from barriers}

The width of massive barriers for all subsequent simulations was fixed
to three peptide groups. The placement of the barrier was such that
the free $\alpha$-helix on one side is twice as long as the free
$\alpha$-helix on the other side, in particular, the barrier spanned
peptide groups $n=13-15$ or $n=26-28$ depending on whether the soliton
is launched from the C-end or the N-end of the helix, respectively.

\begin{figure}
\begin{centering}
\includegraphics[width=160mm]{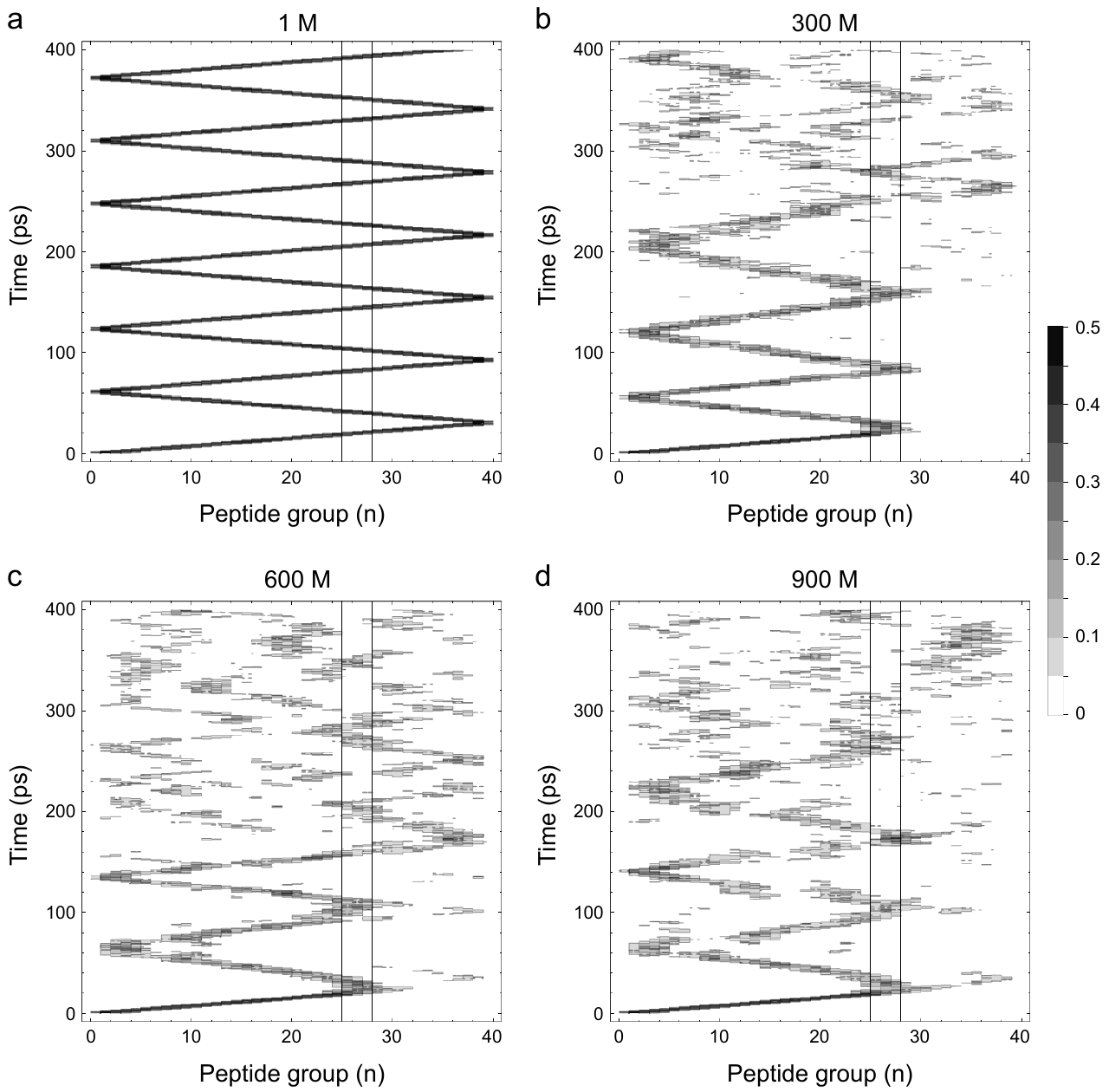}
\par\end{centering}
\caption{\label{fig:2}Soliton reflection from massive barriers extending over
three peptide groups $n=26-28$, each of which with effective mass of
$1M$ (no barrier), $300M$, $600M$ or $900M$, for the isotropic exciton-lattice coupling $\xi=1$ launched by a Gaussian of amide~I energy $\sigma=3$ applied at the
N-end of an $\alpha$-helix spine composed of 40 peptide groups during
a period of 400 ps. The barrier location is indicated with thin vertical
lines.}
\end{figure}

\begin{figure}
\begin{centering}
\includegraphics[width=160mm]{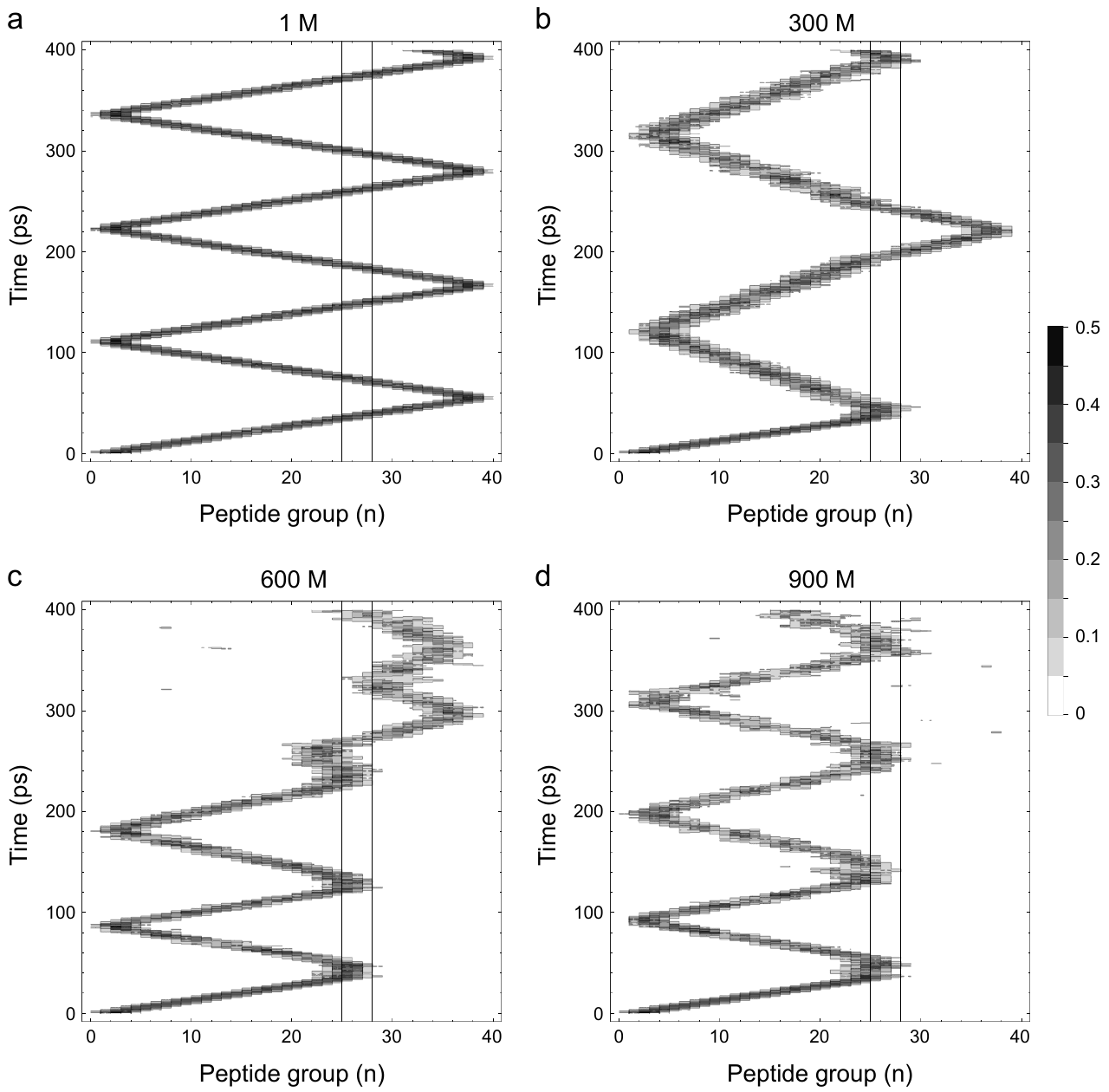}
\par\end{centering}
\caption{\label{fig:3}Soliton reflection from massive barriers extending over
three peptide groups $n=26-28$, each of which with effective mass of
$1M$ (no barrier), $300M$, $600M$ or $900M$, for the isotropic exciton-lattice coupling $\xi=1$ launched by a Gaussian of amide~I energy $\sigma=5$ applied at the
N-end of an $\alpha$-helix spine composed of 40 peptide groups during
a period of 400 ps. The barrier location is indicated with thin vertical
lines.}
\end{figure}

\begin{figure}
\begin{centering}
\includegraphics[width=160mm]{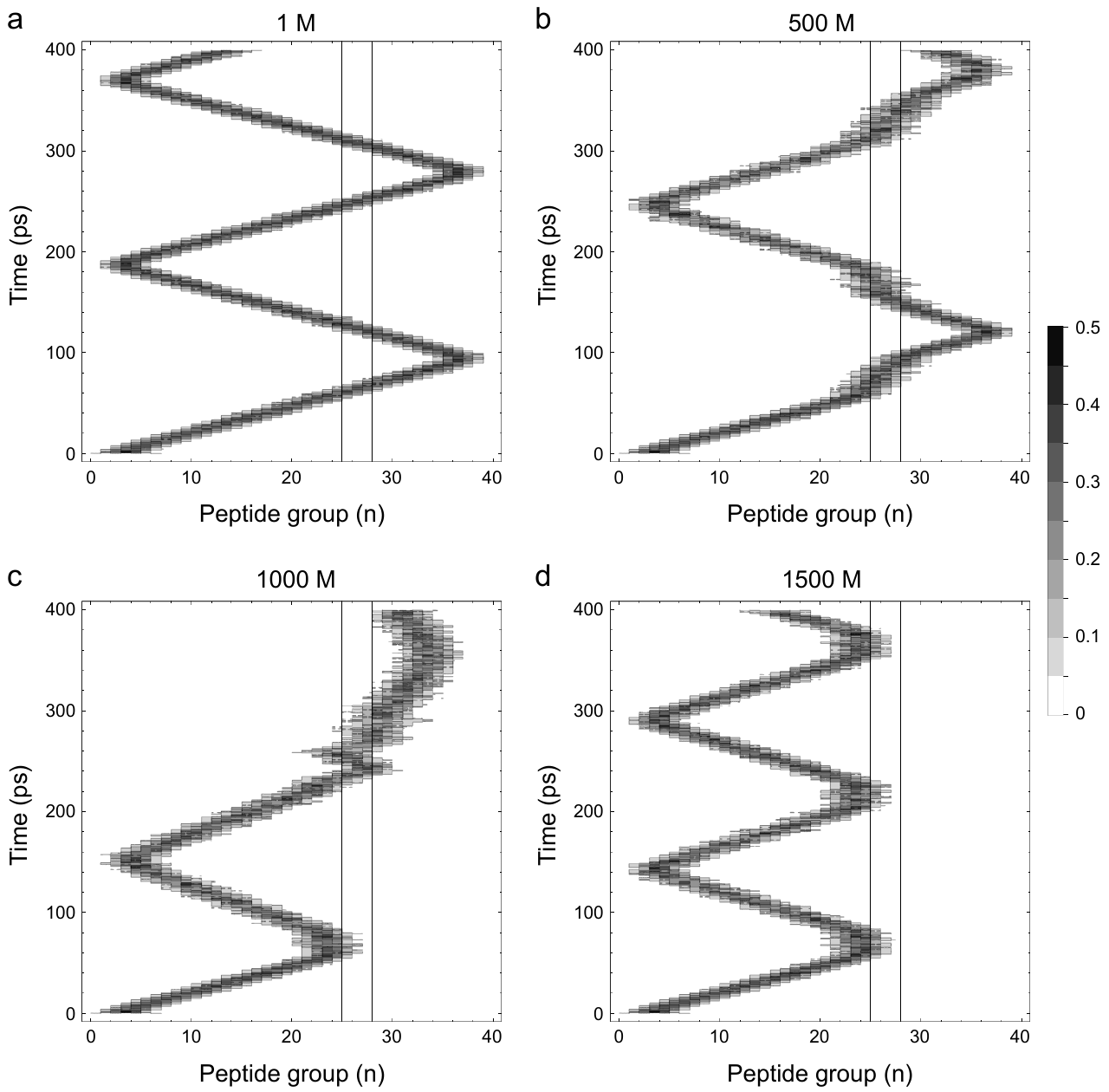}
\par\end{centering}
\caption{\label{fig:4}Soliton reflection from massive barriers extending over
three peptide groups $n=26-28$, each of which with effective mass of
$1M$ (no barrier), $500M$, $1000M$ or $1500M$, for the isotropic exciton-lattice coupling $\xi=1$ launched by a Gaussian of amide~I energy $\sigma=7$ applied at the
N-end of an $\alpha$-helix spine composed of 40 peptide groups during
a period of 400 ps. The barrier location is indicated with thin vertical
lines.}
\end{figure}

Simulations with isotropic exciton-lattice coupling 
$\xi=1$ for a wide range ($1-1500$~M) of effective peptide group
masses inside a rectangular barrier (all peptide groups inside the
barrier having equal mass), showed that the greater width $\sigma$ of
the Davydov soliton stabilizes the soliton upon reflection from the
barrier and prevents disintegration (Figs. \ref{fig:2}, \ref{fig:3}
and \ref{fig:4}). The soliton with width $\sigma=3$ is able to bounce
a couple of times from massive barriers below 600~M before it disintegrates
at a timescale of 100~ps (Fig.~\ref{fig:2}). The soliton with width
$\sigma=5$ is stable in the presence of massive barriers below 900~M
and the soliton can either bounce off the barrier or tunnel through
the barrier (Fig.~\ref{fig:3}a--d, Videos 1--4). The number of bounces off the barrier
before successful tunneling increases with the mass of the barrier.
As an example, for 300~M barrier the soliton with width $\sigma=5$,
bounces once before tunneling (Fig.~\ref{fig:3}b, Video 2), whereas for 600~M
barrier the soliton bounces twice before tunneling (Fig.~\ref{fig:3}c, Video 3).
Conversely, the number of bounces off the barrier before successful
tunneling decreases when the width of the soliton increases (Fig.~\ref{fig:4}).
A further example is the soliton with width $\sigma=7$ that bounces once before
tunneling through 1000 M barrier (Fig.~\ref{fig:4}c), whereas the
soliton with width $\sigma=5$, bounces twice before tunneling through
a much lower 600 M barrier (Fig.~\ref{fig:3}c). Thus, these
results demonstrate that the soliton width increases stability upon
reflection, increases probability for tunneling and delays dispersion.
The results also suggest that the energy released by ATP hydrolysis in biological
systems should be advantageously delivered as a Gaussian pulse to
several amide~I groups (e.g. $\sigma=5$), rather than concentrated
into a single amide~I exciton ($\sigma=1$).

Previous theoretical research utilizing the continuum approximation \cite{Davydov1976,Davydov1979,Davydov1982,Davydov1986,Davydov1988,Scott1984,Scott1985,Scott1992} reduced the discrete system of the Davydov equations to a non-linear Schr\"{o}dinger equation (NLSE), which admits $\textrm{sech}^2$-shaped soliton solutions that can be obtained with the inverse scattering transform and allied techniques in perturbation theory for nearly integrable systems \cite{Zakharov1972,Kivshar1989}. Because the soliton width determines the energy of Davydov solitons, the tunneling processes depend essentially on the ratio between the energy of the tunneling `particle' (soliton) and energetic characteristics of the potential barrier. In nonlinear tunneling processes, the `tails' of solitons play important role (cf.~\cite{Davydov1987,Ermakov1988}), and this is also confirmed by our numerical results of the discrete system of the Davydov equations. In fact, our numerical results validate the continuum approximation used for deriving the NLSE in the case of biologically realistic, short protein $\alpha$-helices, whereas analytical study of the NLSE in turn provides a qualitative explanation of the observed nonlinear phenomena in the presented numerical simulations.
In particular, studies of NLSE with impurities confirm that the greater soliton width is able to promote the tunneling \cite{Forinash1994,Ostrovskaya1999}, whereas the time evolution of different initial excitations for the concrete values of parameters in the Davydov NLSE model of one-dimensional molecular chains, confirms that initial excitations that better resemble the $\textrm{sech}^2$ soliton shape exhibit a lower threshold for soliton formation, i.e. a Gaussian pulse enacted upon several amide~I groups is expected to be more advantageous for generating solitons in comparison with the excitation of a single amide~I group \cite{Brizhik1983,Brizhik1988,Brizhik1993}.

\subsection{Effect of anisotropy of the exciton-lattice coupling on tunneling time}

\begin{figure}
\begin{centering}
\includegraphics[width=160mm]{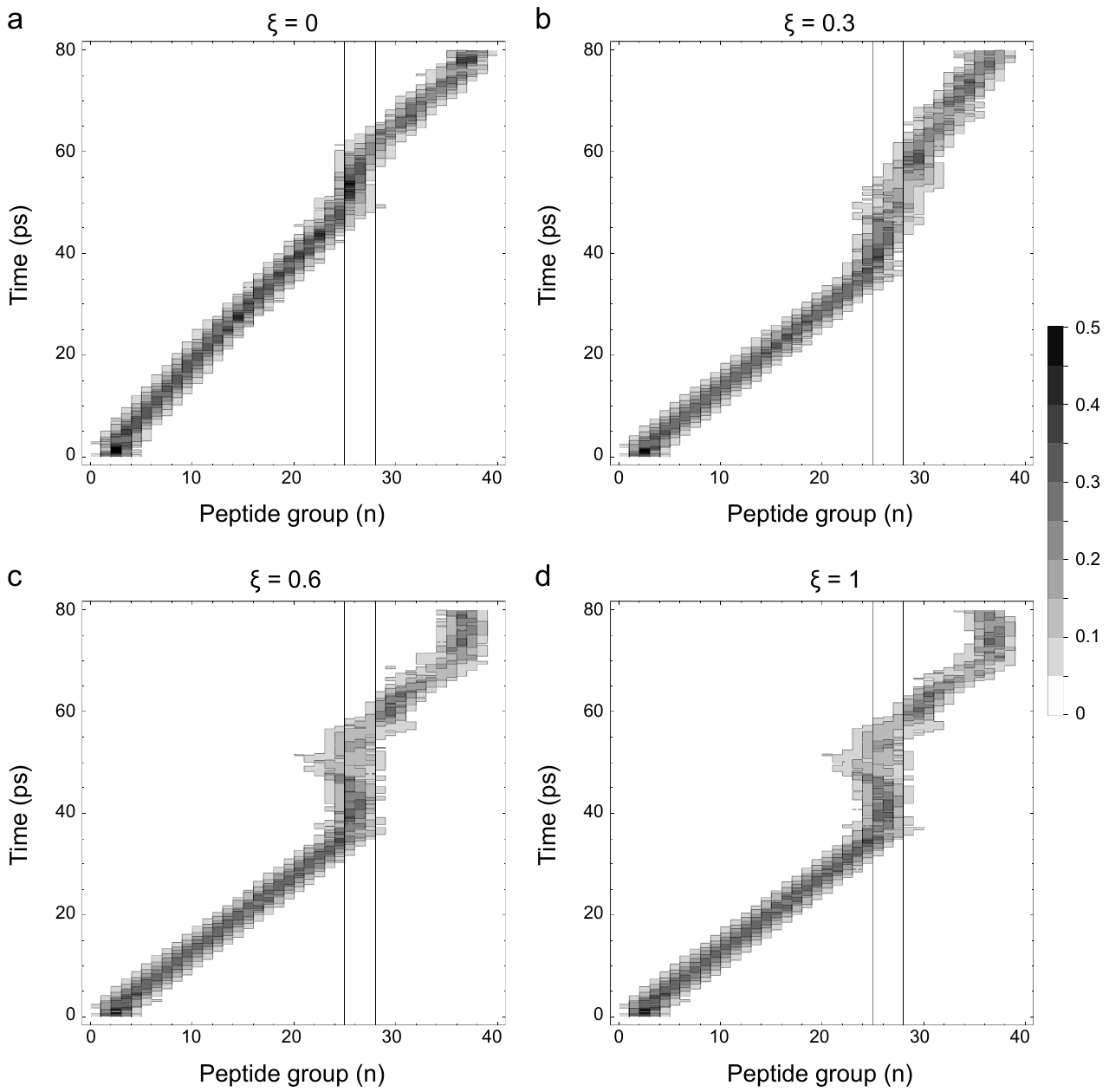}
\par\end{centering}
\caption{\label{fig:5}Soliton tunneling through a massive barrier extending
over three peptide groups $n=26-28$, each of which with effective mass
of $200M$, for different values of the isotropy parameter $\xi=\{0,0.3,0.6,1\}$
launched by a Gaussian of amide~I energy $\sigma=5$ applied at the N-end of an $\alpha$-helix
spine composed of 40 peptide groups during a period of 80 ps. The
barrier location is indicated with thin vertical lines.}
\end{figure}

\begin{figure}
\begin{centering}
\includegraphics[width=160mm]{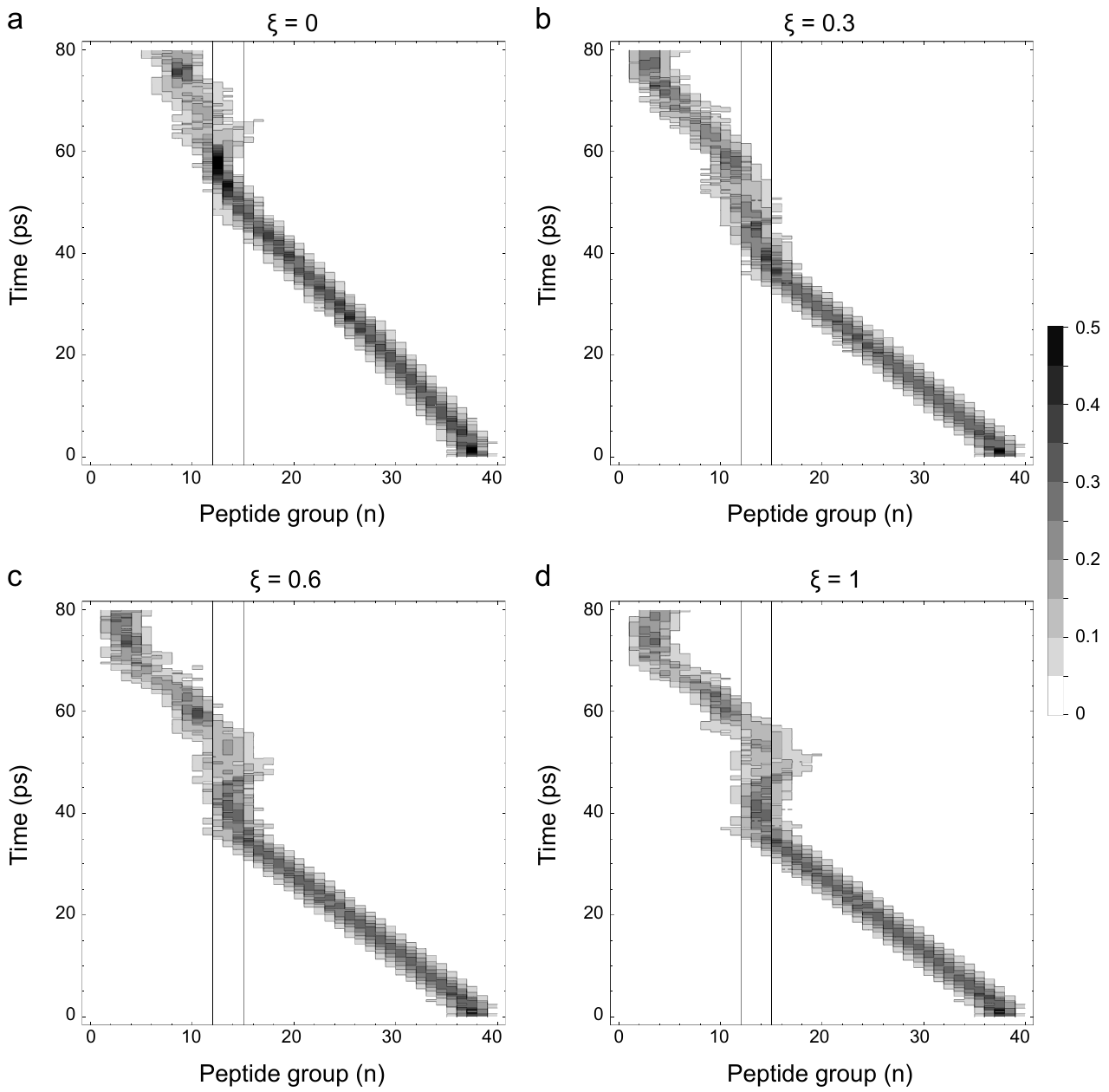}
\par\end{centering}
\caption{\label{fig:6}Soliton tunneling through a massive barrier extending
over three peptide groups $n=13-15$, each of which with effective mass
of $200M$, for different values of the isotropy parameter $\xi=\{0,0.3,0.6,1\}$
launched by a Gaussian of amide~I energy $\sigma=5$ applied at the C-end of an $\alpha$-helix
spine composed of 40 peptide groups during a period of 80 ps. The
barrier location is indicated with thin vertical lines.}
\end{figure}

Full isotropy of the exciton-lattice coupling $\xi=1$
makes the soliton dynamics mirror symmetric in respect to launching
from the N-end or the C-end of the $\alpha$-helix. When the isotropy 
is partial, $\xi<1$, the mirror symmetry of the soliton dynamics is
violated. Therefore, in order to be exhaustive in our analysis of anisotropic exciton-lattice coupling, we
have simulated soliton tunneling through a massive barrier from the
left and from the right (Figs. \ref{fig:5} and \ref{fig:6}). The
higher value for the isotropy parameter $\xi$ prolongs significantly
the tunneling time for both left and right crossing of the barrier.
As an example, the soliton with width $\sigma=5$ tunnels through
200 M barrier from left to right for 16 ps when $\xi=0$ (Fig.~\ref{fig:5}a)
and 24.5 ps when $\xi=1$ (Fig.~\ref{fig:5}d). Because the higher
value of $\xi$ increases the soliton speed, the soliton launched
from the N-end of the $\alpha$-helix with $\xi=1$ reaches the barrier
earlier, and then exits on the other side of the barrier earlier than the
soliton with $\xi=0$, despite the longer tunneling time required
when $\xi=1$. In the case of anisotropic exciton-lattice coupling $\xi=0$, the soliton speed
is slightly faster when moving from left to right (216~m/s, Fig.~\ref{fig:5}a),
as opposed to moving from right to left (214~m/s, Fig.~\ref{fig:6}a).
The tunneling from right to left of solitons launched from the C-end
of the $\alpha$-helix is qualitatively similar, but when $\xi=0$
the soliton widens to a greater extent at the exit of the other side
of the massive barrier (Fig.~\ref{fig:6}a) in comparison to the case
of tunneling from left to right (Fig.~\ref{fig:5}a). Thus, even though
the anisotropy of the exciton-lattice coupling speeds
up the tunneling times, the overall transport of energy through massive
barriers could be delayed in comparison to the fully isotropic case.

\subsection{Effect of anisotropy of the exciton-lattice coupling on tunneling probability}

\begin{figure}
\begin{centering}
\includegraphics[width=160mm]{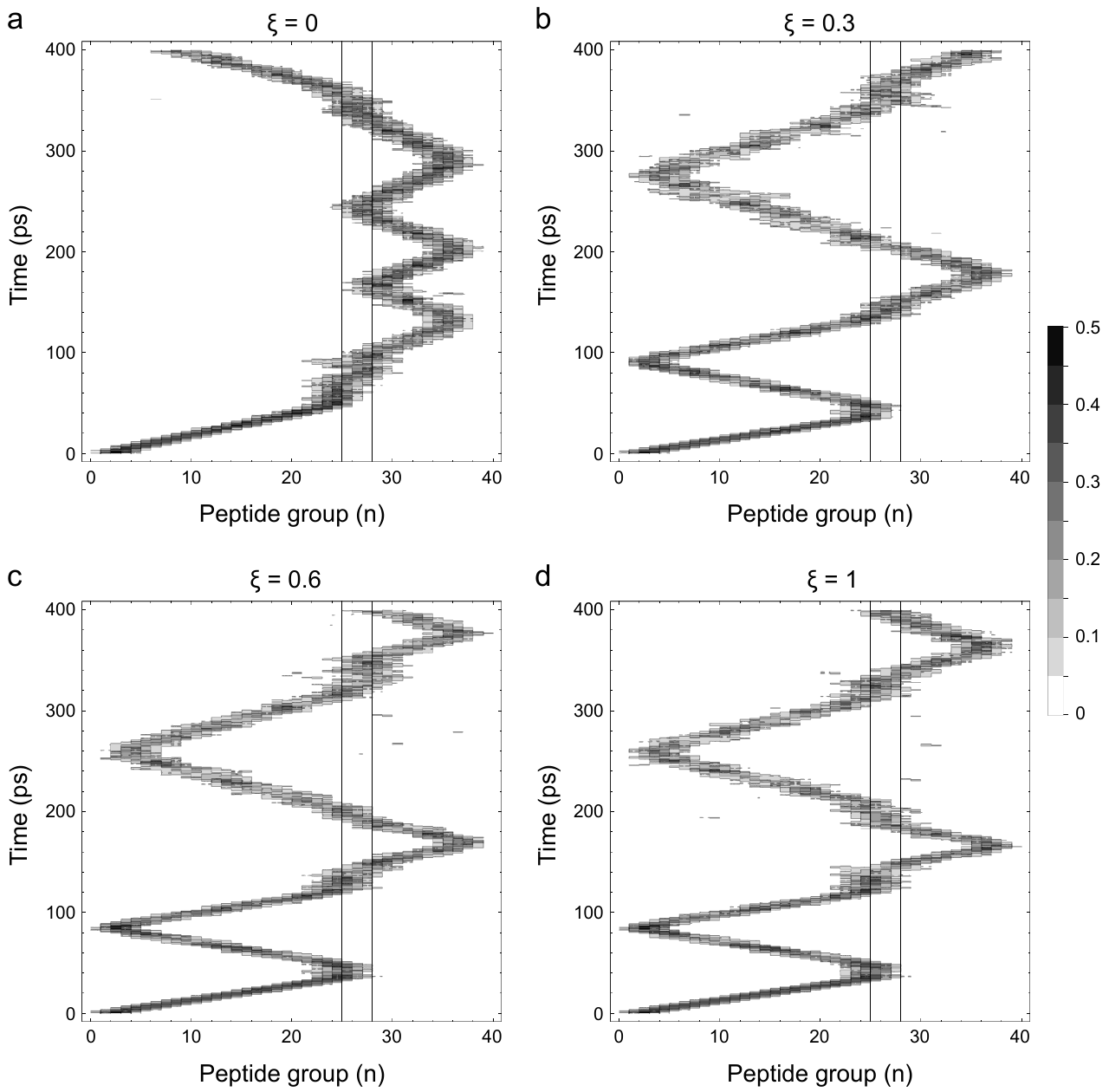}
\par\end{centering}
\caption{\label{fig:7}Soliton tunneling through a massive barrier extending
over three peptide groups $n=26-28$, each of which with effective mass
of $400M$, for different values of the isotropy parameter $\xi=\{0,0.3,0.6,1\}$
launched by a Gaussian of amide~I energy $\sigma=5$ applied at the N-end of an $\alpha$-helix
spine composed of 40 peptide groups during a period of 400 ps. The
barrier location is indicated with thin vertical lines.}
\end{figure}

\begin{figure}
\begin{centering}
\includegraphics[width=160mm]{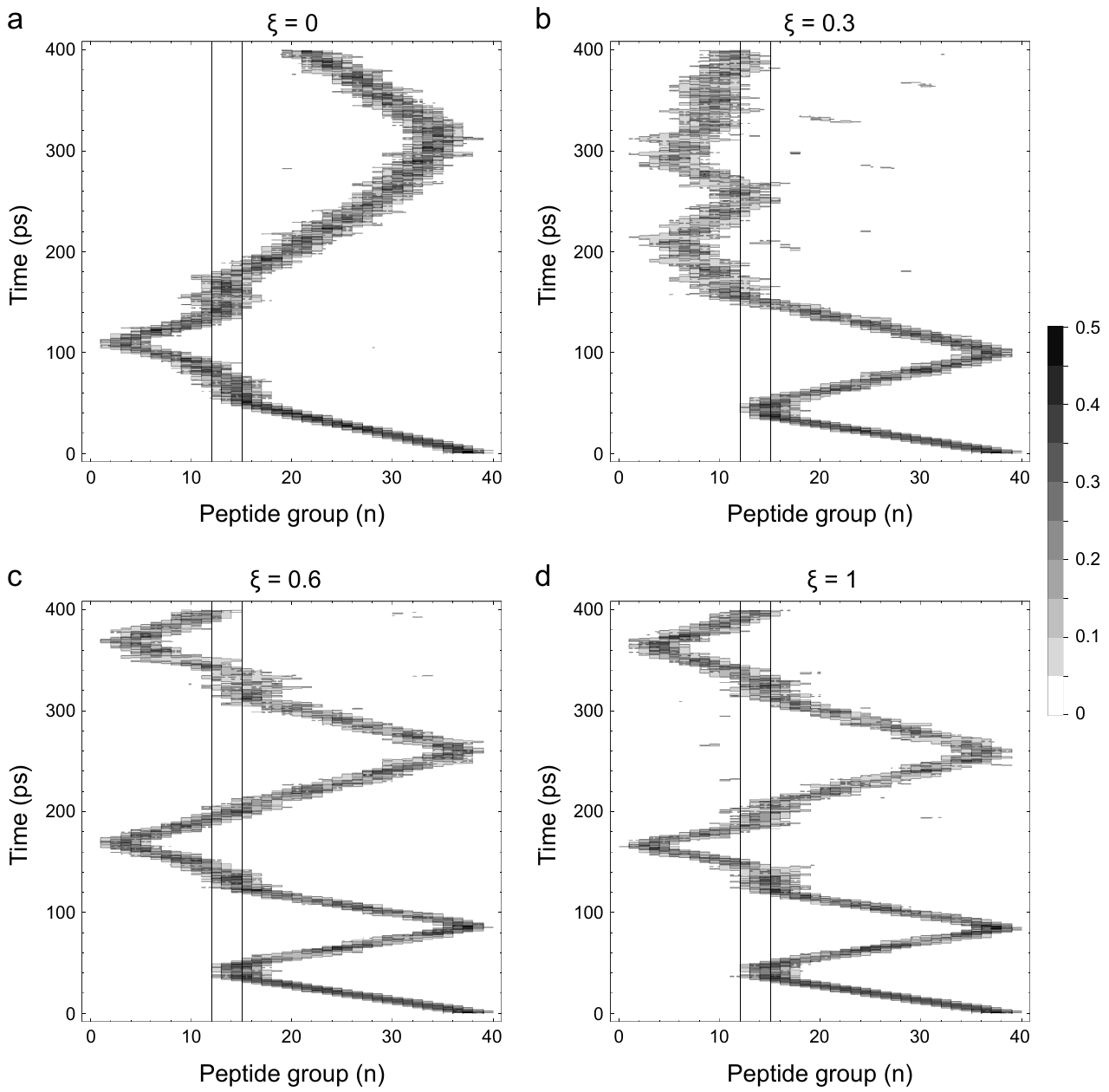}
\par\end{centering}
\caption{\label{fig:8}Soliton tunneling through a massive barrier extending
over three peptide groups $n=13-15$, each of which with effective mass
of $400M$, for different values of the isotropy parameter $\xi=\{0,0.3,0.6,1\}$
launched by a Gaussian of amide~I energy $\sigma=5$ applied at the C-end of an $\alpha$-helix
spine composed of 40 peptide groups during a period of 400 ps. The
barrier location is indicated with thin vertical lines.}
\end{figure}

The overall effect of the isotropy parameter $\xi$ on the transport
of energy across massive barriers includes the tunneling probability
as an additional factor that operates together with the tunneling
time and the soliton speed. From simulations with sufficiently massive
barriers that are capable of reflecting the soliton, it was determined
that higher values of $\xi$ decrease the probability of soliton tunneling,
and increase the probability of reflection. As an example, the soliton
with width $\sigma=5$ tunnels through 400 M barrier without reflection
when $\xi=0$ (Figs.~\ref{fig:7}a and \ref{fig:8}a), whereas it
reflects once before tunneling through the barrier when $\xi=1$ (Figs.~\ref{fig:7}d and \ref{fig:8}d).
Thus, for very massive barriers the anisotropy of the exciton-lattice coupling could speed up
the overall transport of energy through the barrier due to increased probability of tunneling in comparison to the fully isotropic case.

\subsection{Effect of anisotropy of the exciton-lattice coupling on soliton stability}

\begin{figure}
\begin{centering}
\includegraphics[width=160mm]{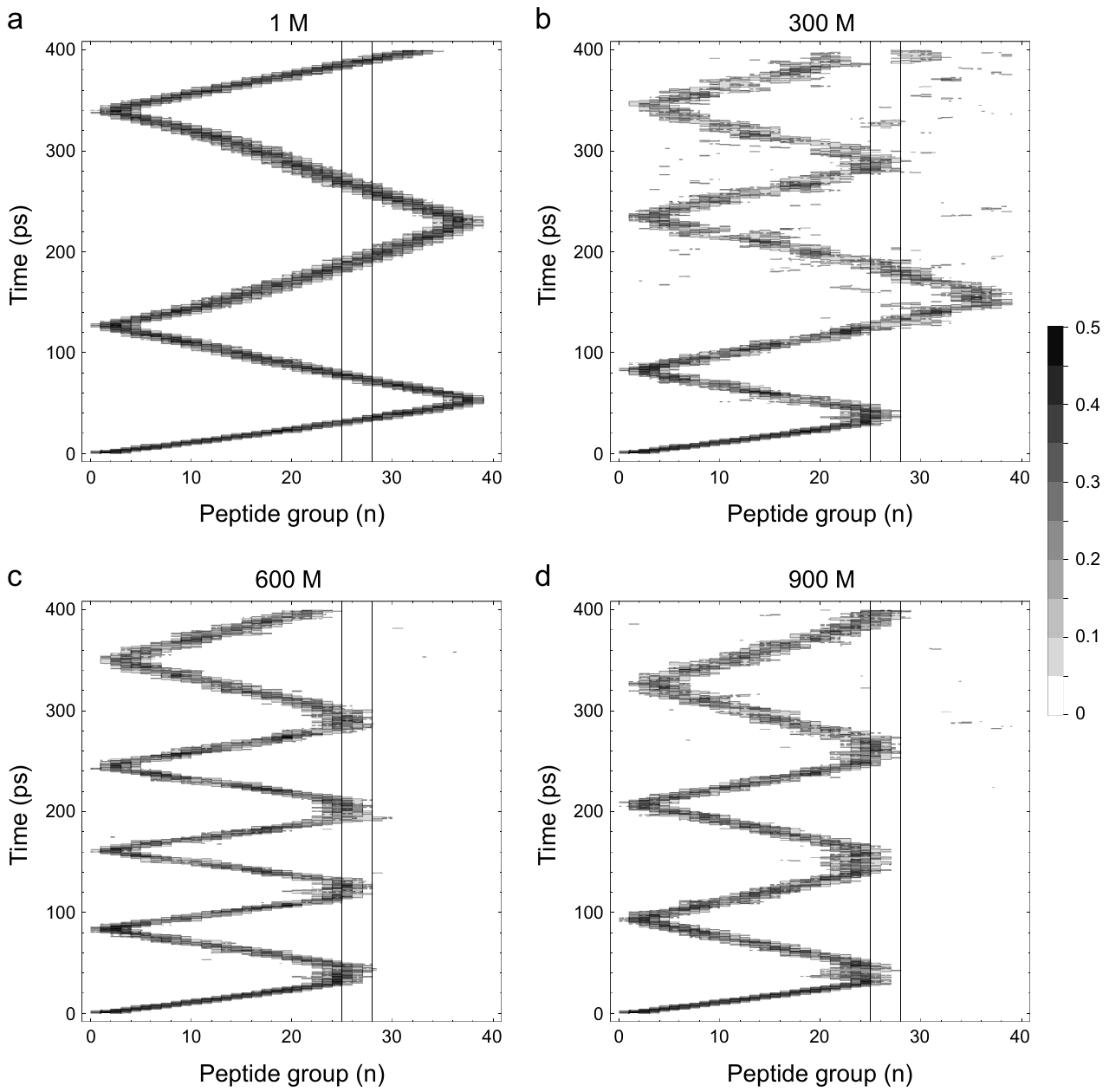}
\par\end{centering}
\caption{\label{fig:9}Soliton reflection from massive barriers extending over
three peptide groups $n=26-28$, each of which with effective mass of
$1M$ (no barrier), $300M$, $600M$ or $900M$, for the anisotropic exciton-lattice coupling $\xi=0$ launched by a Gaussian of amide~I energy $\sigma=3$ applied at the
N-end of an $\alpha$-helix spine composed of 40 peptide groups during
a period of 400 ps. The barrier location is indicated with thin vertical
lines.}
\end{figure}

\begin{figure}
\begin{centering}
\includegraphics[width=160mm]{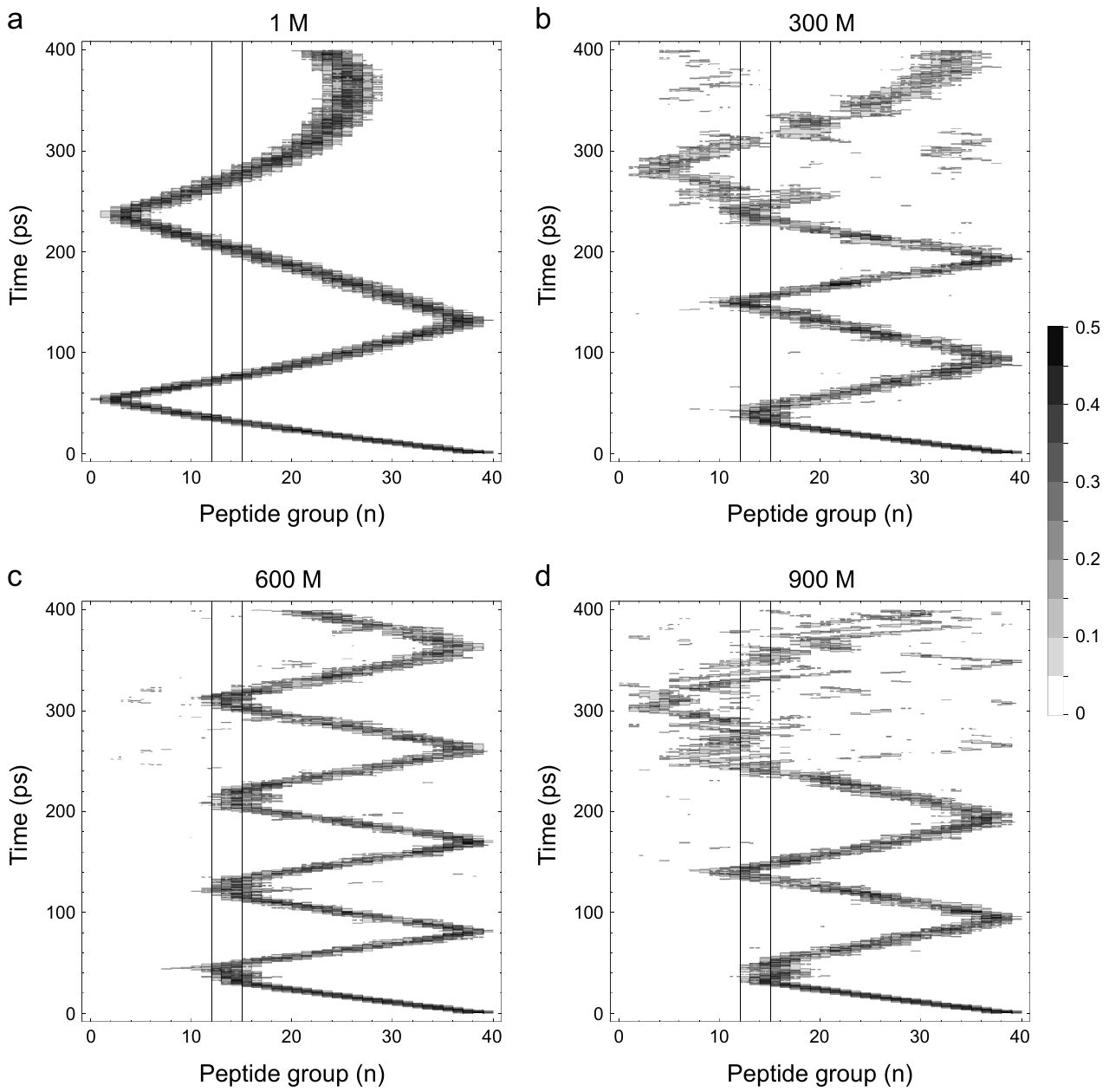}
\par\end{centering}
\caption{\label{fig:10}Soliton reflection from massive barriers extending
over three peptide groups $n=13-15$, each of which with effective mass
of $1M$ (no barrier), $300M$, $600M$ or $900M$, for the anisotropic exciton-lattice coupling $\xi=0$ launched by a Gaussian of amide~I energy $\sigma=3$ applied at the
C-end of an $\alpha$-helix spine composed of 40 peptide groups during
a period of 400 ps. The barrier location is indicated with thin vertical
lines.}
\end{figure}

The higher values of the isotropy parameter $\xi$ lead to decreased
stability of the solitons upon reflection from massive barriers. Comparison
of simulations with $\xi=1$ (the soliton dynamics is mirror symmetric
for launching from the N-end versus launching from the C-end of the
$\alpha$-helix) or $\xi=0$ (the soliton dynamics is not mirror symmetric),
shows that for sufficiently massive barriers, repetitive bouncing of the soliton
leads to soliton destabilization and dispersion more easily when the isotropy parameter
$\xi$ has higher values. As an example, the soliton with width $\sigma=3$ impacting
upon a 900 M barrier for $\xi=1$, bounces once before its dispersion (Fig.~\ref{fig:2}d),
whereas for $\xi=0$, it bounces twice on the right of
the barrier (Fig.~\ref{fig:10}d), or bounces thrice on the left of
the barrier (Fig.~\ref{fig:9}d). Thus, the presence of structural
 anisotropy in the chain of hydrogen bonded peptide groups along the
$\alpha$-helix spine
\[
\cdots H-N-C=O\cdots H-N-C=O\cdots H-N-C=O\cdots
\]
which leads to stronger coupling of the amide~I exciton ($C=O$ group)
to the hydrogen bond on the right ($-C=O\cdots H$) as compared to
the hydrogen bond on the left ($O\cdots H-N-C=O$), namely, $\chi_{r}>\chi_{l}$
\cite{Kuprievich1990}, could effectively enhance the stability of solitons bouncing
off massive barriers as created by the presence of external protein clamps
acting on the protein $\alpha$-helix. The dependence of the soliton
stability on the direction of barrier impact, also shows that protein
$\alpha$-helices with anisotropic exciton-lattice coupling ($\xi<1$)
may then have a preferential direction for transmission of energy inside
proteins.

\begin{figure}
\begin{centering}
\includegraphics[width=160mm]{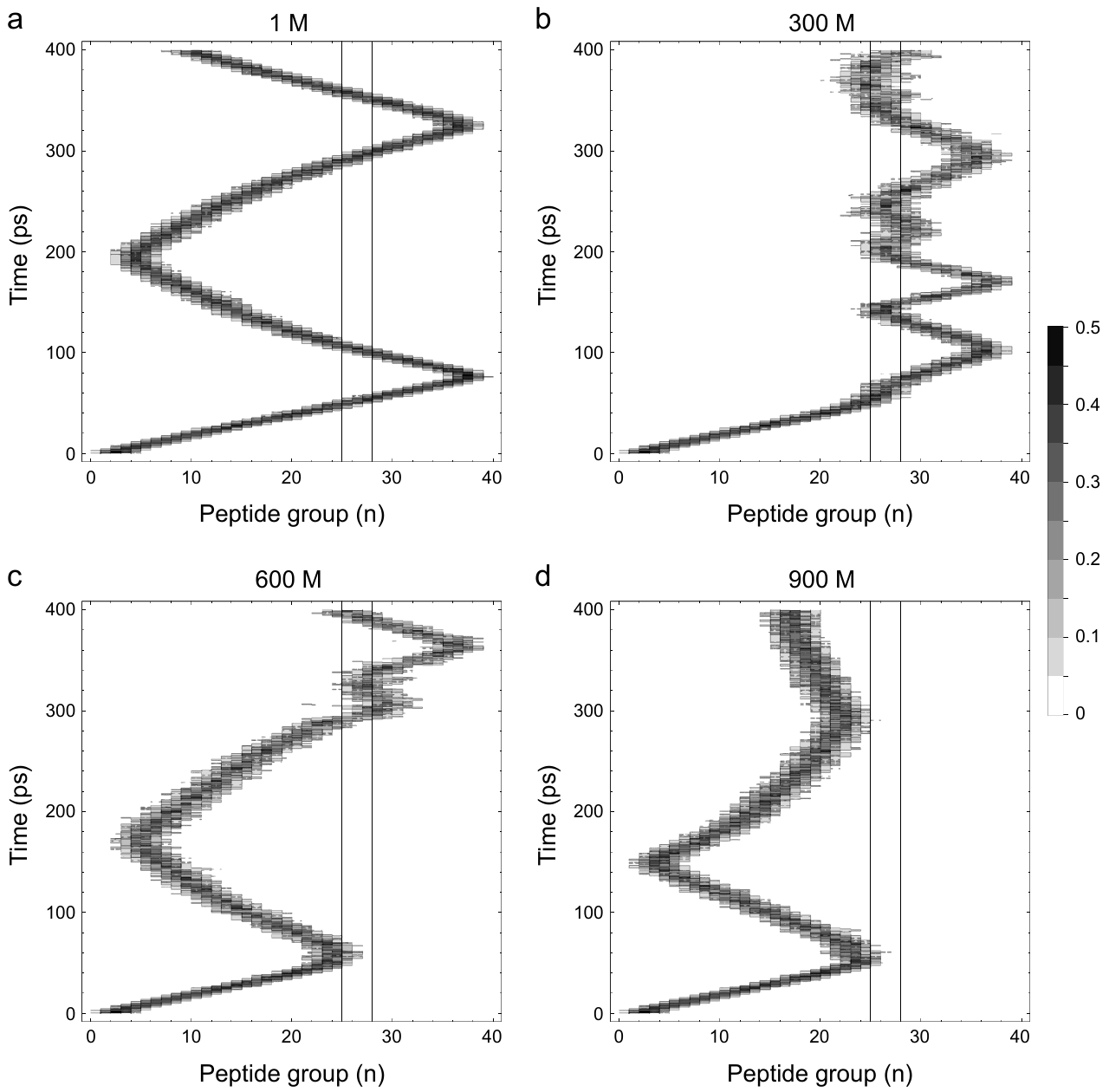}
\par\end{centering}
\caption{\label{fig:11}Soliton reflection from massive barriers extending
over three peptide groups $n=26-28$, each of which with effective mass
of $1M$ (no barrier), $300M$, $600M$ or $900M$, for the anisotropic exciton-lattice coupling $\xi=0$ launched by a Gaussian of amide~I energy $\sigma=5$ applied at the
N-end of an $\alpha$-helix spine composed of 40 peptide groups during
a period of 400 ps. The barrier location is indicated with thin vertical
lines.}
\end{figure}

\begin{figure}
\begin{centering}
\includegraphics[width=160mm]{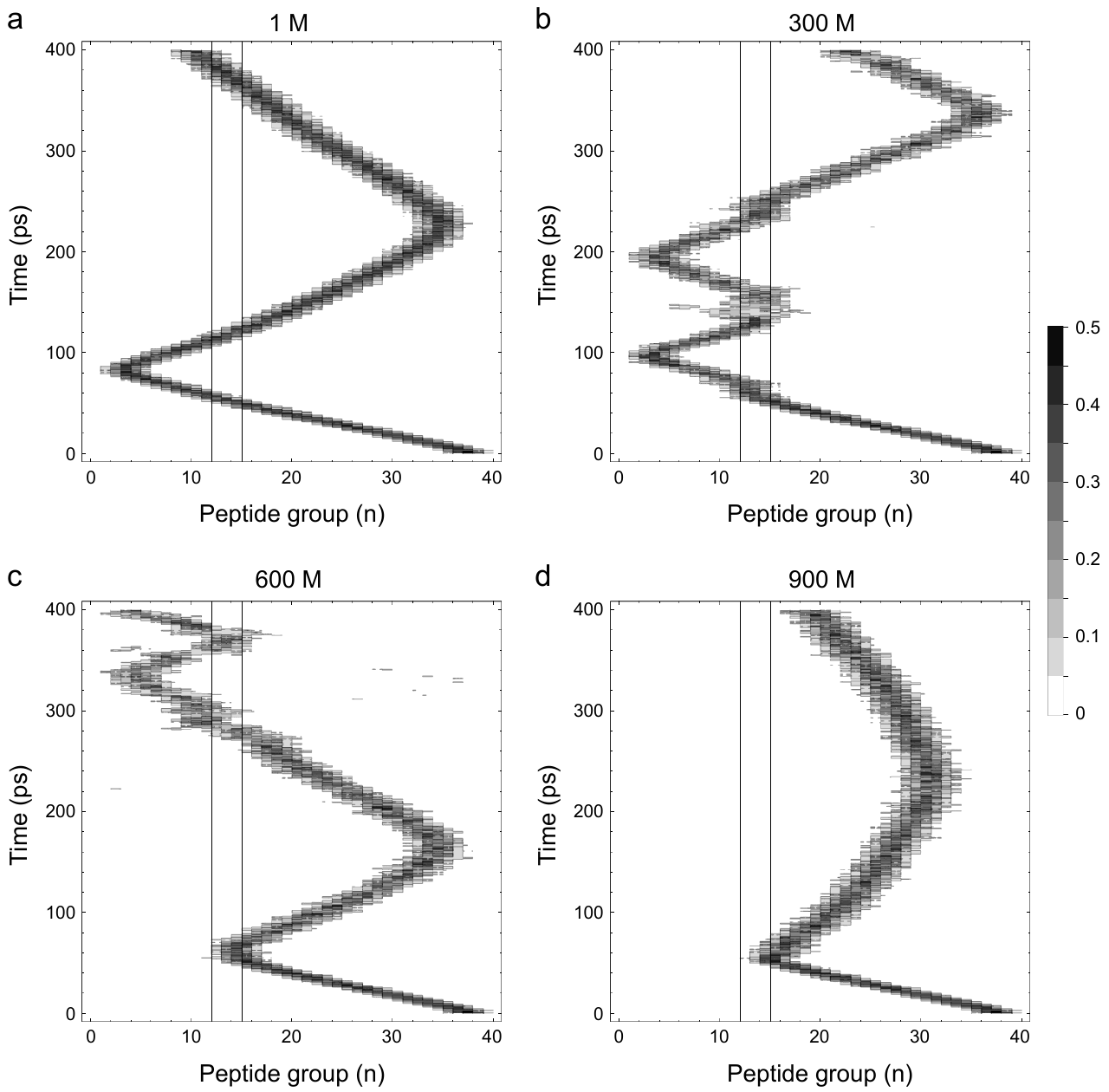}
\par\end{centering}
\caption{\label{fig:12}Soliton reflection from massive barriers extending
over three peptide groups $n=13-15$, each of which with effective mass
of $1M$ (no barrier), $300M$, $600M$ or $900M$, for the anisotropic exciton-lattice coupling $\xi=0$ launched by a Gaussian of amide~I energy $\sigma=5$ applied at the
C-end of an $\alpha$-helix spine composed of 40 peptide groups during
a period of 400 ps. The barrier location is indicated with thin vertical
lines.}
\end{figure}

\begin{figure}
\begin{centering}
\includegraphics[width=160mm]{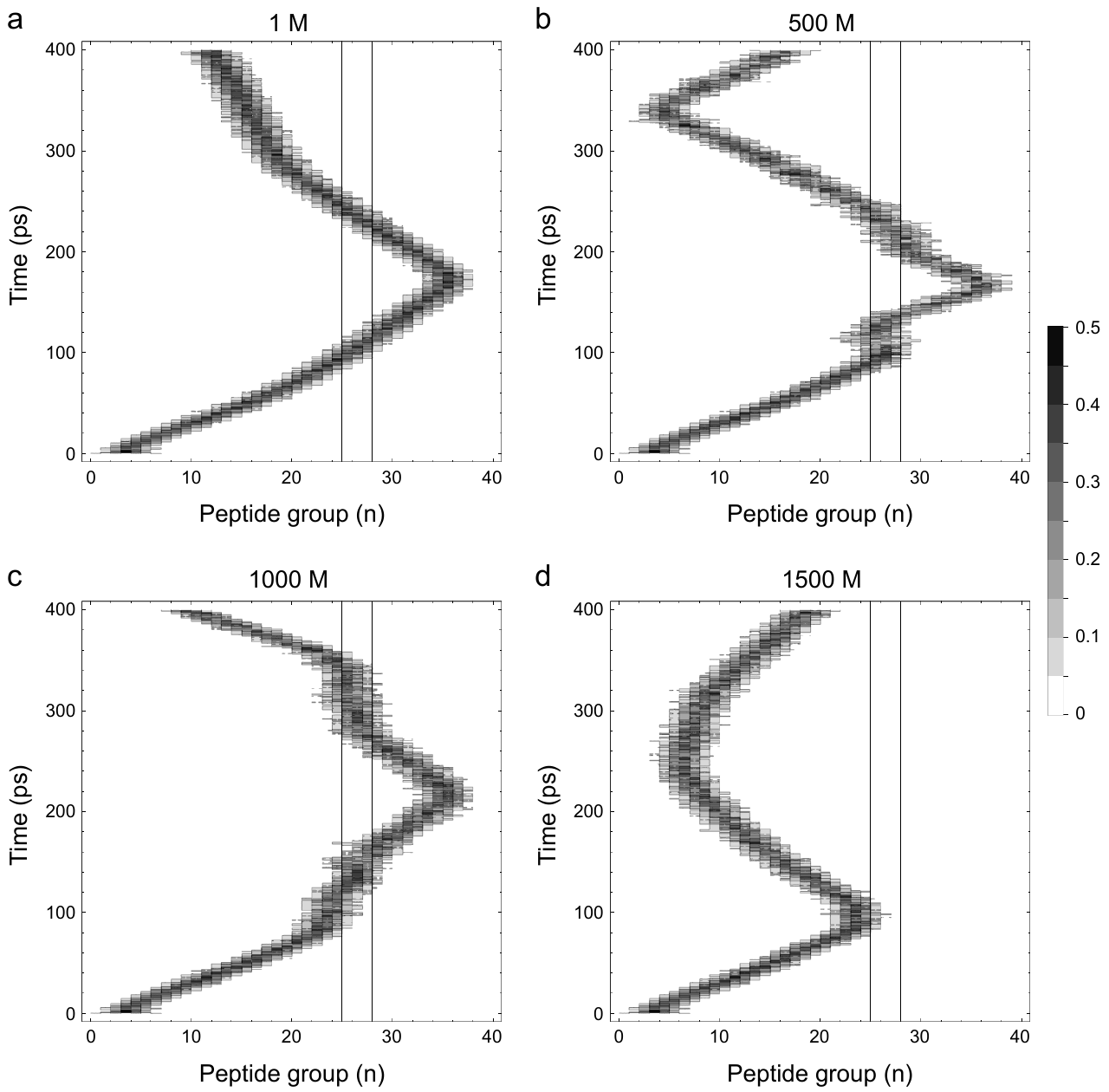}
\par\end{centering}
\caption{\label{fig:13}Soliton reflection from massive barriers extending
over three peptide groups $n=26-28$, each of which with effective mass
of $1M$ (no barrier), $300M$, $600M$ or $900M$, for the anisotropic exciton-lattice coupling $\xi=0$ launched by a Gaussian of amide~I energy $\sigma=7$ applied at the N-end of an $\alpha$-helix spine
composed of 40 peptide groups during a period of 400 ps.
The barrier location is indicated with thin vertical lines.}
\end{figure}

\begin{figure}
\begin{centering}
\includegraphics[width=160mm]{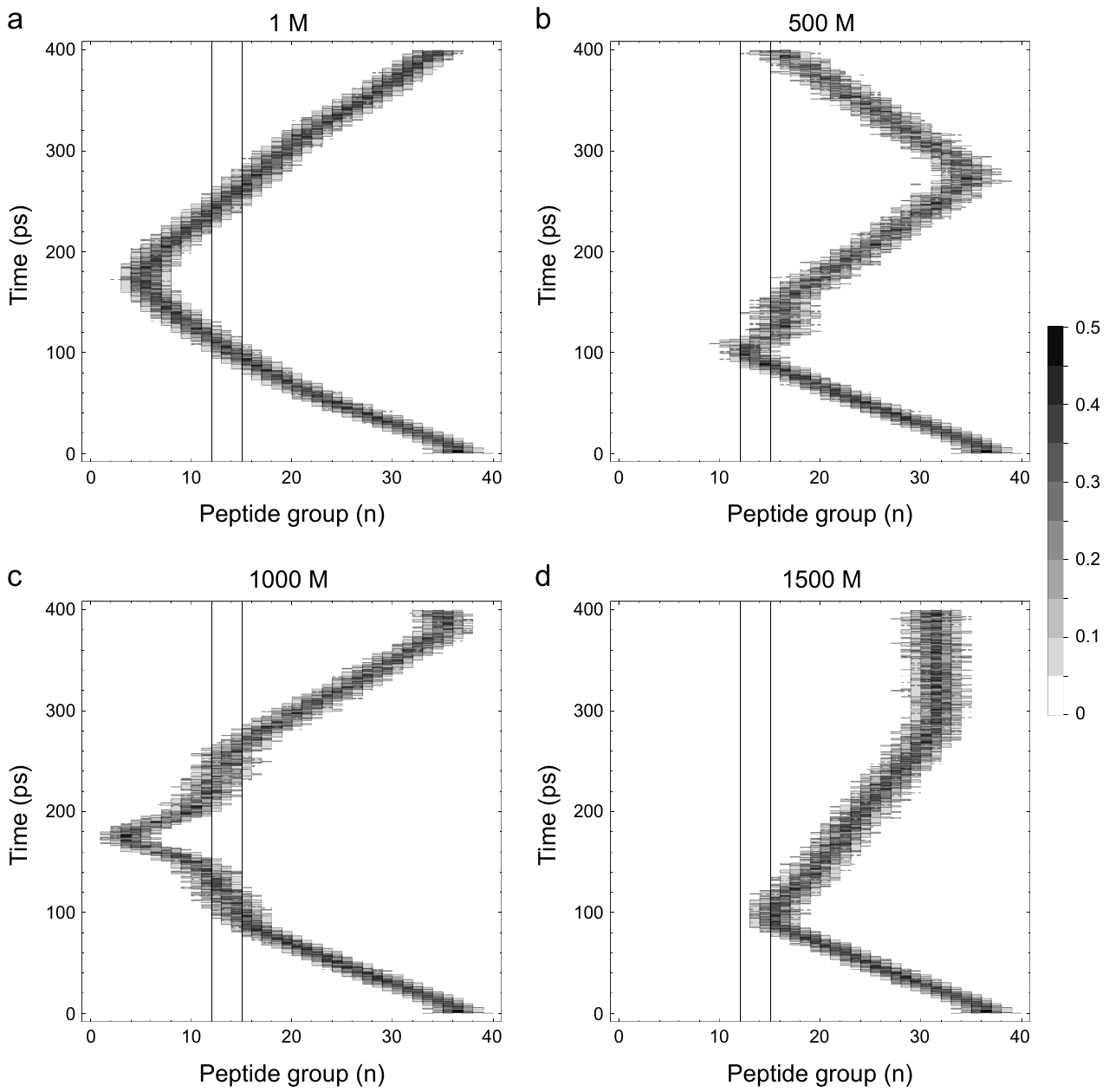}
\par\end{centering}
\caption{\label{fig:14}Soliton reflection from massive barriers extending
over three peptide groups $n=13-15$, each of which with effective mass
of $1M$ (no barrier), $300M$, $600M$ or $900M$, for the anisotropic exciton-lattice coupling $\xi=0$ launched by a Gaussian of amide~I energy $\sigma=7$ applied at the C-end of an $\alpha$-helix spine
composed of 40 peptide groups during a period of 400 ps.
The barrier location is indicated with thin vertical lines.}
\end{figure}

Simulations with wider solitons $\sigma=5$ (Figs. \ref{fig:3}, \ref{fig:11}
and \ref{fig:12}) or $\sigma=7$ (Figs. \ref{fig:4}, \ref{fig:13}
and \ref{fig:14}) corroborate the findings that the higher
$\xi$ increases both the soliton instability upon barrier impact
and the probability for reflection from the massive barrier.
This further highlights the need of improved \emph{ab initio} quantum chemical calculations for
determining of the exact value of the isotropy parameter $\xi=\frac{\chi_{l}}{\chi_{r}}$.

\subsection{Launching of solitons from massive barriers}

\begin{figure}
\begin{centering}
\includegraphics[width=160mm]{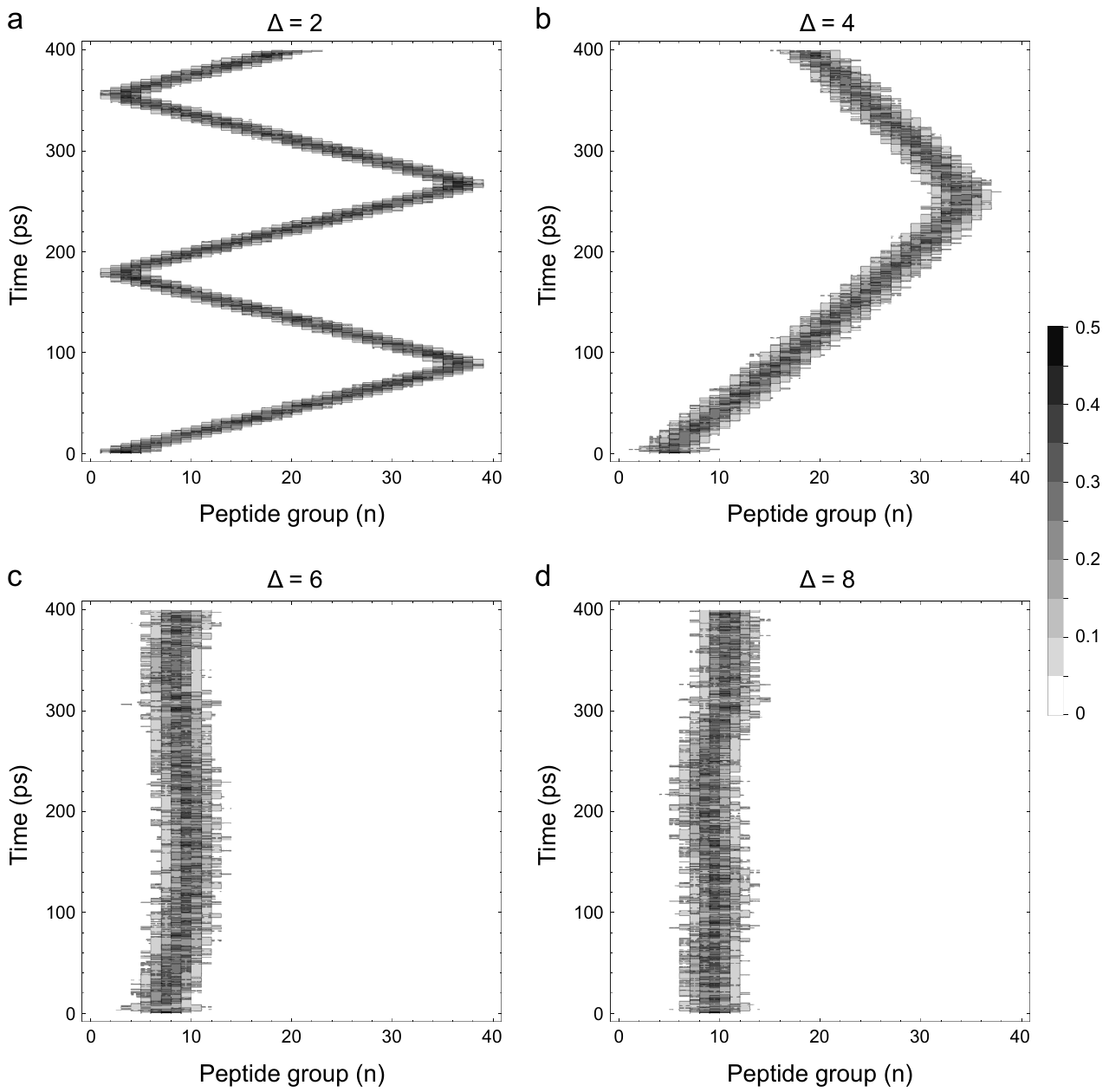}
\par\end{centering}
\caption{\label{fig:15}Launching of soliton by a Gaussian of amide~I energy $\sigma=5$ applied
at a distance $\Delta=\{2,4,6,8\}$ peptide groups away from the N-end
of an $\alpha$-helix spine composed of 40 peptide groups for the isotropic exciton-lattice coupling $\xi=1$ during a period of 400 ps.}
\end{figure}

\begin{figure}
\begin{centering}
\includegraphics[width=160mm]{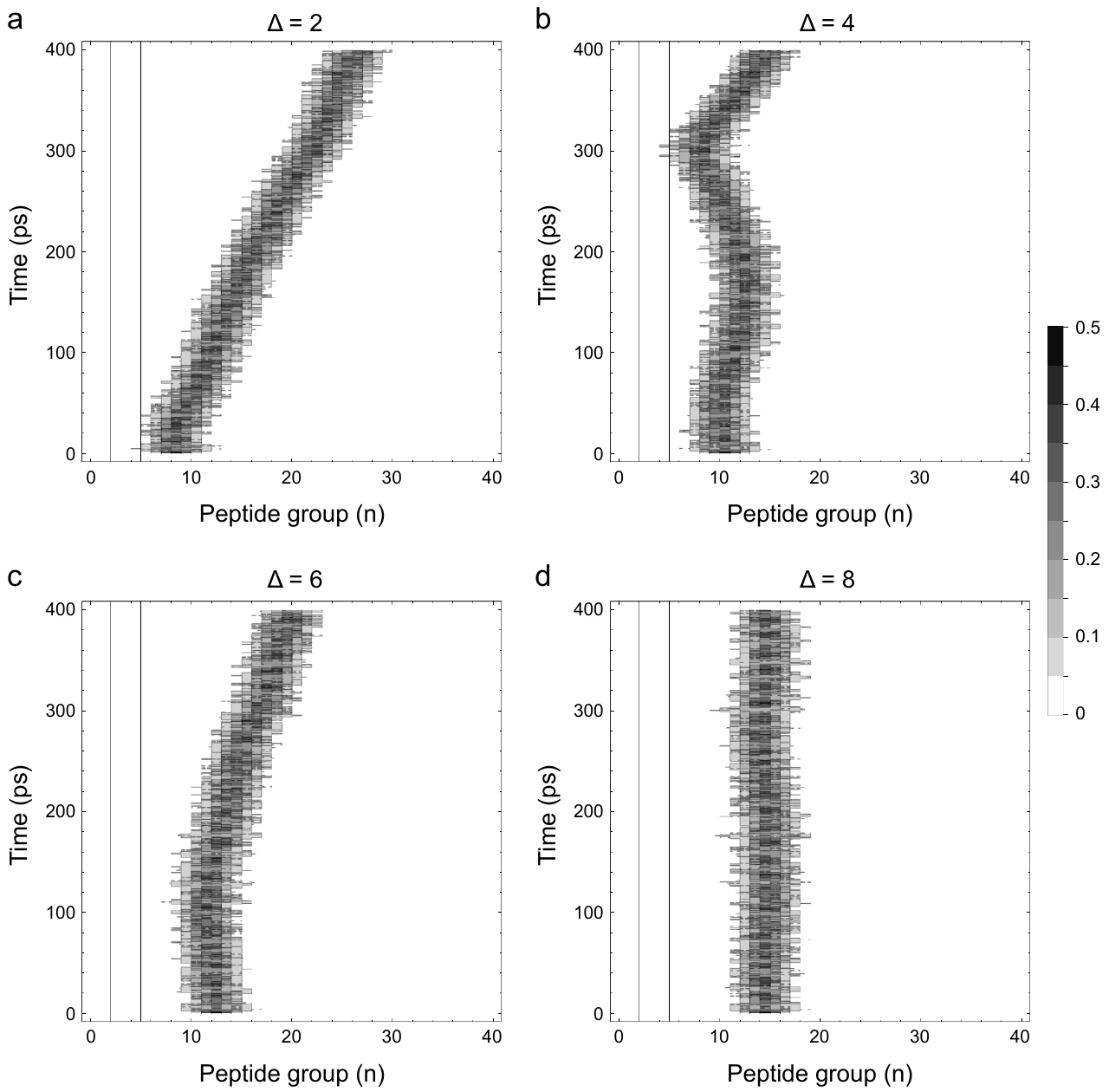}
\par\end{centering}
\caption{\label{fig:16}Launching of soliton by a Gaussian of amide~I energy $\sigma=5$ applied
at a distance $\Delta=\{2,4,6,8\}$ peptide groups away from a massive
barrier extending over three peptide groups $n=3-5$, each of which with
effective mass of $500M$, of an $\alpha$-helix spine composed of
40 peptide groups for the isotropic exciton-lattice coupling $\xi=1$ during a
period of 400 ps. The barrier location is indicated with thin vertical
lines.}
\end{figure}

\begin{figure}
\begin{centering}
\includegraphics[width=160mm]{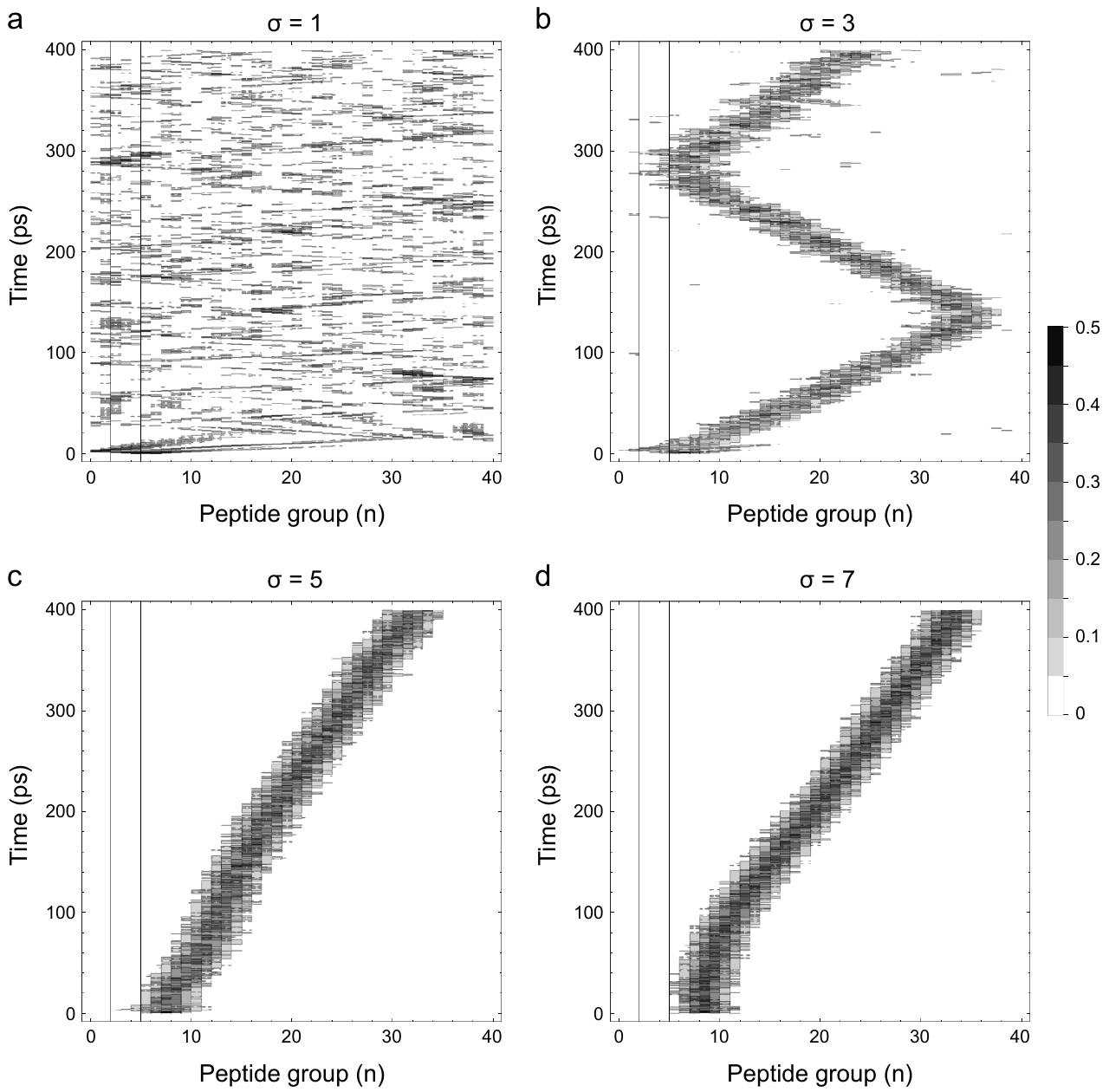}
\par\end{centering}
\caption{\label{fig:17}Launching of solitons by different Gaussians of amide~I energy $\sigma=\{1,3,5,7\}$
applied at peptide groups adjacent to a massive barrier extending
over three peptide groups $n=3-5$, each of which with effective mass
of $500M$, of an $\alpha$-helix spine composed of 40 peptide groups
for the isotropic exciton-lattice coupling $\xi=1$ during a period of 400 ps.
The barrier location is indicated with thin vertical lines.}
\end{figure}

The finite size of protein $\alpha$-helices introduces reflective
boundary conditions that enable launching of Davydov solitons from
either end of the $\alpha$-helix \cite{GeorgievGlazebrook2019}.
Simulations with $\xi=1$ of solitons with different width~$\sigma$
showed that propagating solitons are launched only if the amide~I energy
is within several peptide groups from one of the ends of the $\alpha$-helix.
As an example, a Gaussian of amide~I energy with width $\sigma=5$ produces
propagating solitons when applied at a distance $\Delta\leq4$ peptide
groups from the $\alpha$-helix end, and pinned solitons when~$\Delta>4$
(Fig.~\ref{fig:15}). To test whether massive barriers can also be
used for launching propagating solitons, we have simulated a massive 500
M barrier located at peptide groups $n=3-5$, and applied a Gaussian of amide~I energy at a position where it generates
a pinned soliton
in the absence of the barrier (Fig.~\ref{fig:15}c). Indeed, a propagating
soliton was launched (Fig.~\ref{fig:16}a), albeit moving with a substantially
lower speed in comparison with the case when it is launched
at the $\alpha$-helix end (Fig.~\ref{fig:15}a). We also investigated the efficiency of
soliton launching for different Gaussians of amide~I energy $\sigma=\{1,3,5,7\}$, as applied to peptide groups adjacent
to the massive barrier. The simulations showed that the soliton is
immediately dispersed if the amide~I exciton energy is applied to
a single peptide group (Fig.~\ref{fig:17}a), but launches a propagating
soliton for $\sigma\geq3$ (Figs. \ref{fig:17}b, c, d).
Thus, protein clamps could be able to launch propagating Davydov solitons when energy is supplied by
biochemical processes that trigger the clamp removal.

\section{Conclusions}

Following works such as \cite{Devault1984}, the possibility of quantum mechanical tunneling effects in biological systems has for about 30 years been a significant focus of attention (surveyed in \cite{Brookes2017}). Evidence for such effects has arisen through detection of such phenomena as the kinetic isotope effect in enzymatic reactions \cite{Klinman2013}, and vibrationally assisted tunneling for classes of proteins
\cite{Basran2001,Sutcliffe2000} (see also \cite{GeorgievGlazebrook2018} which underscores the role of amide I vibrational energy). In close connection with this scope of research, we have shown in this present paper that
the action of external protein clamps could be physically modeled by a local increment of the effective mass of peptide groups inside a clamped protein $\alpha$-helix, and that such clamping action introduces a massive barrier that could either reflect Davydov solitons, or allow them to tunnel through.
We have also identified several factors that affect the transmission of energy
by Davydov solitons in the presence of such massive barriers, namely, the soliton speed,
the tunneling probability, and the tunneling time. In line with a previous
quantum model of SNARE protein zipping in active zones of neuronal synapses \cite{GeorgievGlazebrook2018},
the simulations, as presented, demonstrate that Davydov solitons are indeed capable
of bouncing off a massive barrier multiple times before successfully
tunneling into the other side of the barrier.
These results may have implications for an improved understanding of protein-protein interactions, synaptic function, and the transfer of information between neurons in neural networks.

\section*{Conflict of interest}

The authors declare that they have no conflict of interest.

\section*{Acknowledgment}

We would like to thank an anonymous reviewer for providing useful suggestions for improving the presentation of this work.

\appendix
\section{Supplementary videos}

\paragraph{Video~1}
Soliton dynamics in the absence of a massive barrier for the case shown in Fig.~\ref{fig:3}a with isotropic exciton-lattice coupling $\xi=1$ launched by a Gaussian of amide~I energy $\sigma=5$ applied at the N-end of an $\alpha$-helix spine composed of 40 peptide groups (extending along the $x$-axis) during a period of 400 ps. Quantum probabilities $|a_n|^2$ are plotted in blue along the $z$-axis. Phonon lattice displacement differences $b_n-b_{n-1}$ (measured in picometers) are plotted in red along the $y$-axis.

\paragraph{Video~2}
Soliton reflection from and tunneling through a massive barrier extending over three peptide groups $n=26-28$, each of which with effective mass of $300M$, for the case shown in Fig.~\ref{fig:3}b with isotropic exciton-lattice coupling $\xi=1$ launched by a Gaussian of amide~I energy $\sigma=5$ applied at the N-end of an $\alpha$-helix spine composed of 40 peptide groups (extending along the $x$-axis) during a period of 400 ps. Quantum probabilities $|a_n|^2$ are plotted in blue along the $z$-axis. Phonon lattice displacement differences $b_n-b_{n-1}$ (measured in picometers) are plotted in red along the $y$-axis. The place of the massive barrier is indicated with two parallel lines.

\paragraph{Video~3}
Soliton reflection from and tunneling through a massive barrier extending over three peptide groups $n=26-28$, each of which with effective mass of $600M$, for the case shown in Fig.~\ref{fig:3}c with isotropic exciton-lattice coupling $\xi=1$ launched by a Gaussian of amide~I energy $\sigma=5$ applied at the N-end of an $\alpha$-helix spine composed of 40 peptide groups (extending along the $x$-axis) during a period of 400 ps. Quantum probabilities $|a_n|^2$ are plotted in blue along the $z$-axis. Phonon lattice displacement differences $b_n-b_{n-1}$ (measured in picometers) are plotted in red along the $y$-axis. The place of the massive barrier is indicated with two parallel lines.

\paragraph{Video~4}
Soliton reflection from a massive barrier extending over three peptide groups $n=26-28$, each of which with effective mass of $900M$, for the case shown in Fig.~\ref{fig:3}d with isotropic exciton-lattice coupling $\xi=1$ launched by a Gaussian of amide~I energy $\sigma=5$ applied at the N-end of an $\alpha$-helix spine composed of 40 peptide groups (extending along the $x$-axis) during a period of 400 ps. Quantum probabilities $|a_n|^2$ are plotted in blue along the $z$-axis. Phonon lattice displacement differences $b_n-b_{n-1}$ (measured in picometers) are plotted in red along the $y$-axis. The place of the massive barrier is indicated with two parallel lines.

\bibliography{references}

\begin{thebibliography}{10}
\expandafter\ifx\csname url\endcsname\relax
  \def\url#1{\texttt{#1}}\fi
\expandafter\ifx\csname urlprefix\endcsname\relax\def\urlprefix{URL }\fi
\expandafter\ifx\csname href\endcsname\relax
  \def\href#1#2{#2} \def\path#1{#1}\fi

\bibitem{Strong2004}
M.~Strong, Protein nanomachines, PLoS Biology 2~(3) (2004) e73.
\newblock \href {https://doi.org/10.1371/journal.pbio.0020073}
  {\path{doi:10.1371/journal.pbio.0020073}}.

\bibitem{Hess2005}
H.~Hess, G.~D. Bachand, Biomolecular motors, Materials Today 8~(12, Supplement
  1) (2005) 22--29.
\newblock \href {https://doi.org/10.1016/S1369-7021(05)71286-4}
  {\path{doi:10.1016/S1369-7021(05)71286-4}}.

\bibitem{Hirokawa2010}
N.~Hirokawa, S.~Niwa, Y.~Tanaka, Molecular motors in neurons: transport
  mechanisms and roles in brain function, development, and disease, Neuron
  68~(4) (2010) 610--638.
\newblock \href {https://doi.org/10.1016/j.neuron.2010.09.039}
  {\path{doi:10.1016/j.neuron.2010.09.039}}.

\bibitem{Ha2016}
T.~Ha, Probing nature's nanomachines one molecule at a time, Biophysical
  Journal 110~(5) (2016) 1004--1007.
\newblock \href {https://doi.org/10.1016/j.bpj.2016.02.009}
  {\path{doi:10.1016/j.bpj.2016.02.009}}.

\bibitem{Jones1996}
S.~Jones, J.~M. Thornton, Principles of protein-protein interactions,
  Proceedings of the National Academy of Sciences of the United States of
  America 93~(1) (1996) 13--20.
\newblock \href {https://doi.org/10.1073/pnas.93.1.13}
  {\path{doi:10.1073/pnas.93.1.13}}.

\bibitem{Ottman2013}
C.~Ottman, Protein-protein interactions: an overview, in: A.~D\"{o}mling (Ed.),
  Protein-Protein Interactions in Drug Discovery, Methods and Principles in
  Medicinal Chemistry, Wiley-VCH, Weinheim, Germany, 2013, pp. 1--19.
\newblock \href {https://doi.org/10.1002/9783527648207.ch1}
  {\path{doi:10.1002/9783527648207.ch1}}.

\bibitem{Keskin2016}
O.~Keskin, N.~Tuncbag, A.~Gursoy, Predicting protein-protein interactions from
  the molecular to the proteome level, Chemical Reviews 116~(8) (2016)
  4884--4909.
\newblock \href {https://doi.org/10.1021/acs.chemrev.5b00683}
  {\path{doi:10.1021/acs.chemrev.5b00683}}.

\bibitem{Giraudo2006}
C.~G. Giraudo, W.~S. Eng, T.~J. Melia, J.~E. Rothman, A clamping mechanism
  involved in {SNARE}-dependent exocytosis, Science 313~(5787) (2006) 676--680.
\newblock \href {https://doi.org/10.1126/science.1129450}
  {\path{doi:10.1126/science.1129450}}.

\bibitem{GeorgievGlazebrook2007}
D.~D. Georgiev, J.~F. Glazebrook, Subneuronal processing of information by
  solitary waves and stochastic processes, in: S.~E. Lyshevski (Ed.), Nano and
  Molecular Electronics Handbook, CRC Press, Boca Raton, 2007, Ch.~17, pp.
  1--41.

\bibitem{GeorgievGlazebrook2012}
D.~D. Georgiev, J.~F. Glazebrook, Quasiparticle tunneling in neurotransmitter
  release, in: W.~A. Goddard~III, D.~Brenner, S.~E. Lyshevski, G.~J. Iafrate
  (Eds.), Handbook of Nanoscience, Engineering, and Technology, 3rd Edition,
  CRC Press, Boca Raton, 2012, Ch.~30, pp. 983--1016.

\bibitem{Georgiev2017}
D.~D. Georgiev, Quantum Information and Consciousness: A Gentle Introduction,
  CRC Press, Boca Raton, 2017.
\newblock \href {https://doi.org/10.1201/9780203732519}
  {\path{doi:10.1201/9780203732519}}.

\bibitem{Li2013}
C.~Li, M.~Enomoto, A.~M. Rossi, M.-D. Seo, T.~Rahman, P.~B. Stathopulos, C.~W.
  Taylor, M.~Ikura, J.~B. Ames, {CaBP1}, a neuronal {C}a$^{2+}$ sensor protein,
  inhibits inositol trisphosphate receptors by clamping intersubunit
  interactions, Proceedings of the National Academy of Sciences of the United
  States of America 110~(21) (2013) 8507--8512.
\newblock \href {https://doi.org/10.1073/pnas.1220847110}
  {\path{doi:10.1073/pnas.1220847110}}.

\bibitem{Simms2014}
B.~A. Simms, G.~W. Zamponi, Neuronal voltage-gated calcium channels: structure,
  function, and dysfunction, Neuron 82~(1) (2014) 24--45.
\newblock \href {https://doi.org/10.1016/j.neuron.2014.03.016}
  {\path{doi:10.1016/j.neuron.2014.03.016}}.

\bibitem{Hashemifar2018}
S.~Hashemifar, B.~Neyshabur, A.~A. Khan, J.~Xu, Predicting protein-protein
  interactions through sequence-based deep learning, Bioinformatics 34~(17)
  (2018) i802--i810.
\newblock \href {https://doi.org/10.1093/bioinformatics/bty573}
  {\path{doi:10.1093/bioinformatics/bty573}}.

\bibitem{Douma2017}
L.~G. Douma, K.~K. Yu, J.~K. England, M.~Levitus, L.~B. Bloom, Mechanism of
  opening a sliding clamp, Nucleic Acids Research 45~(17) (2017) 10178--10189.
\newblock \href {https://doi.org/10.1093/nar/gkx665}
  {\path{doi:10.1093/nar/gkx665}}.

\bibitem{Iyer2013}
J.~Iyer, C.~J. Wahlmark, G.~A. Kuser-Ahnert, F.~Kawasaki, Molecular mechanisms
  of complexin fusion clamp function in synaptic exocytosis revealed in a new
  {D}rosophila mutant, Molecular and Cellular Neuroscience 56 (2013) 244--254.
\newblock \href {https://doi.org/10.1016/j.mcn.2013.06.002}
  {\path{doi:10.1016/j.mcn.2013.06.002}}.

\bibitem{Davydov1976}
A.~S. Davydov, N.~I. Kislukha, Solitons in one-dimensional molecular chains,
  Physica Status Solidi (b) 75~(2) (1976) 735--742.
\newblock \href {https://doi.org/10.1002/pssb.2220750238}
  {\path{doi:10.1002/pssb.2220750238}}.

\bibitem{Davydov1979}
A.~S. Davydov, Solitons in molecular systems, Physica Scripta 20~(3-4) (1979)
  387--394.
\newblock \href {https://doi.org/10.1088/0031-8949/20/3-4/013}
  {\path{doi:10.1088/0031-8949/20/3-4/013}}.

\bibitem{Davydov1982}
A.~S. Davydov, Solitons in quasi-one-dimensional molecular structures, Soviet
  Physics Uspekhi 25~(12) (1982) 898--918.
\newblock \href {https://doi.org/10.1070/pu1982v025n12abeh005012}
  {\path{doi:10.1070/pu1982v025n12abeh005012}}.

\bibitem{Davydov1986}
A.~S. Davydov, Quantum theory of the motion of a quasi-particle in a molecular
  chain with thermal vibrations taken into account, Physica Status Solidi (b)
  138~(2) (1986) 559--576.
\newblock \href {https://doi.org/10.1002/pssb.2221380221}
  {\path{doi:10.1002/pssb.2221380221}}.

\bibitem{Davydov1988}
A.~S. Davydov, V.~N. Ermakov, Soliton generation at the boundary of a molecular
  chain, Physica D: Nonlinear Phenomena 32~(2) (1988) 318--323.
\newblock \href {https://doi.org/10.1016/0167-2789(88)90059-0}
  {\path{doi:10.1016/0167-2789(88)90059-0}}.

\bibitem{Scott1984}
A.~C. Scott, Launching a {D}avydov soliton: {I}. {S}oliton analysis, Physica
  Scripta 29~(3) (1984) 279--283.
\newblock \href {https://doi.org/10.1088/0031-8949/29/3/016}
  {\path{doi:10.1088/0031-8949/29/3/016}}.

\bibitem{Scott1985}
A.~C. Scott, Davydov solitons in polypeptides, Philosophical Transactions of
  the Royal Society of London Series A, Mathematical and Physical Sciences
  315~(1533) (1985) 423--436.
\newblock \href {https://doi.org/10.1098/rsta.1985.0049}
  {\path{doi:10.1098/rsta.1985.0049}}.

\bibitem{Scott1992}
A.~C. Scott, Davydov's soliton, Physics Reports 217~(1) (1992) 1--67.
\newblock \href {https://doi.org/10.1016/0370-1573(92)90093-F}
  {\path{doi:10.1016/0370-1573(92)90093-F}}.

\bibitem{Brizhik2004}
L.~S. Brizhik, A.~Eremko, B.~Piette, W.~Zakrzewski, Solitons in
  $\alpha$-helical proteins, Physical Review E 70~(3) (2004) 031914.
\newblock \href {https://doi.org/10.1103/PhysRevE.70.031914}
  {\path{doi:10.1103/PhysRevE.70.031914}}.

\bibitem{Brizhik2006}
L.~Brizhik, A.~Eremko, B.~Piette, W.~Zakrzewski, Charge and energy transfer by
  solitons in low-dimensional nanosystems with helical structure, Chemical
  Physics 324~(1) (2006) 259--266.
\newblock \href {https://doi.org/10.1016/j.chemphys.2006.01.033}
  {\path{doi:10.1016/j.chemphys.2006.01.033}}.

\bibitem{Brizhik2010}
L.~Brizhik, A.~Eremko, B.~Piette, W.~Zakrzewski, Ratchet effect of {D}avydov's
  solitons in nonlinear low-dimensional nanosystems, International Journal of
  Quantum Chemistry 110~(1) (2010) 25--37.
\newblock \href {https://doi.org/10.1002/qua.22083}
  {\path{doi:10.1002/qua.22083}}.

\bibitem{Luo2017}
J.~Luo, B.~Piette, A generalised {D}avydov-{S}cott model for polarons in linear
  peptide chains, European Physical Journal B 90~(8) (2017) 155.
\newblock \href {https://doi.org/10.1140/epjb/e2017-80209-2}
  {\path{doi:10.1140/epjb/e2017-80209-2}}.

\bibitem{GeorgievGlazebrook2019}
D.~D. Georgiev, J.~F. Glazebrook, On the quantum dynamics of {D}avydov solitons
  in protein $\alpha$-helices, Physica A: Statistical Mechanics and its
  Applications 517 (2019) 257--269.
\newblock \href {https://doi.org/10.1016/j.physa.2018.11.026}
  {\path{doi:10.1016/j.physa.2018.11.026}}.

\bibitem{Sun2010}
J.~Sun, B.~Luo, Y.~Zhao, Dynamics of a one-dimensional {H}olstein polaron with
  the {D}avydov ans\"{a}tze, Physical Review B 82~(1) (2010) 014305.
\newblock \href {https://doi.org/10.1103/PhysRevB.82.014305}
  {\path{doi:10.1103/PhysRevB.82.014305}}.

\bibitem{Luo2011}
B.~Luo, J.~Ye, Y.~Zhao, Variational study of polaron dynamics with the
  {D}avydov ans\"{a}tze, Physica Status Solidi (c) 8~(1) (2011) 70--73.
\newblock \href {https://doi.org/10.1002/pssc.201000721}
  {\path{doi:10.1002/pssc.201000721}}.

\bibitem{Zhou2015}
N.~Zhou, Z.~Huang, J.~Zhu, V.~Chernyak, Y.~Zhao, Polaron dynamics with a
  multitude of {D}avydov $d_2$ trial states, Journal of Chemical Physics
  143~(1) (2015) 014113.
\newblock \href {https://doi.org/10.1063/1.4923009}
  {\path{doi:10.1063/1.4923009}}.

\bibitem{Kerr1987}
W.~C. Kerr, P.~S. Lomdahl, Quantum-mechanical derivation of the equations of
  motion for {D}avydov solitons, Physical Review B 35~(7) (1987) 3629--3632.
\newblock \href {https://doi.org/10.1103/PhysRevB.35.3629}
  {\path{doi:10.1103/PhysRevB.35.3629}}.

\bibitem{Kerr1990}
W.~C. Kerr, P.~S. Lomdahl, Quantum-mechanical derivation of the {D}avydov
  equations for multi-quanta states, in: P.~L. Christiansen, A.~C. Scott
  (Eds.), Davydov's Soliton Revisited: Self-Trapping of Vibrational Energy in
  Protein, Springer, New York, 1990, pp. 23--30.
\newblock \href {https://doi.org/10.1007/978-1-4757-9948-4_2}
  {\path{doi:10.1007/978-1-4757-9948-4_2}}.

\bibitem{Itoh1972}
K.~Itoh, T.~Shimanouchi, Vibrational spectra of crystalline formamide, Journal
  of Molecular Spectroscopy 42~(1) (1972) 86--99.
\newblock \href {https://doi.org/10.1016/0022-2852(72)90146-4}
  {\path{doi:10.1016/0022-2852(72)90146-4}}.

\bibitem{MacNeil1984}
L.~MacNeil, A.~C. Scott, Launching a {D}avydov soliton: {II}. {N}umerical
  studies, Physica Scripta 29~(3) (1984) 284--287.
\newblock \href {https://doi.org/10.1088/0031-8949/29/3/017}
  {\path{doi:10.1088/0031-8949/29/3/017}}.

\bibitem{Cruzeiro1988}
L.~Cruzeiro, J.~Halding, P.~L. Christiansen, O.~Skovgaard, A.~C. Scott,
  Temperature effects on the {D}avydov soliton, Physical Review A 37~(3) (1988)
  880--887.
\newblock \href {https://doi.org/10.1103/PhysRevA.37.880}
  {\path{doi:10.1103/PhysRevA.37.880}}.

\bibitem{Petzold1983}
L.~Petzold, Automatic selection of methods for solving stiff and nonstiff
  systems of ordinary differential equations, SIAM Journal on Scientific and
  Statistical Computing 4~(1) (1983) 136--148.
\newblock \href {https://doi.org/10.1137/0904010} {\path{doi:10.1137/0904010}}.

\bibitem{Hindmarsh1983}
A.~C. Hindmarsh, {ODEPACK}, a systematized collection of {ODE} solvers, in:
  R.~S. Stepleman, M.~Carver, R.~Peskin, W.~F. Ames, R.~Vichnevetsky (Eds.),
  Scientific Computing, {IMACS} Transactions on Scientific Computation,
  North-Holland, Amsterdam, 1983, pp. 55--64.

\bibitem{Hindmarsh1995}
A.~C. Hindmarsh, L.~R. Petzold, Algorithms and software for ordinary
  differential equations and differential-algebraic equations, {P}art {II}:
  {H}igher-order methods and software packages, Computers in Physics 9~(2)
  (1995) 148--155.
\newblock \href {https://doi.org/10.1063/1.168540}
  {\path{doi:10.1063/1.168540}}.

\bibitem{Trott2006}
M.~Trott, The Mathematica GuideBook for Numerics, Springer, New York, 2006.
\newblock \href {https://doi.org/10.1007/0-387-28814-7}
  {\path{doi:10.1007/0-387-28814-7}}.

\bibitem{Motschmann1989}
H.~Motschmann, W.~F\"{o}rner, J.~Ladik, Influence of heat bath and disorder in
  the sequence of amino acid masses on {D}avydov solitons, Journal of Physics:
  Condensed Matter 1~(31) (1989) 5083--5093.
\newblock \href {https://doi.org/10.1088/0953-8984/1/31/007}
  {\path{doi:10.1088/0953-8984/1/31/007}}.

\bibitem{Forner1990}
W.~F\"{o}rner, J.~Ladik, Influence of heat bath and disorder on {D}avydov
  solitons, in: P.~L. Christiansen, A.~C. Scott (Eds.), Davydov's Soliton
  Revisited: Self-Trapping of Vibrational Energy in Protein, Springer, New
  York, 1990, pp. 267--283.
\newblock \href {https://doi.org/10.1007/978-1-4757-9948-4_20}
  {\path{doi:10.1007/978-1-4757-9948-4_20}}.

\bibitem{Forner1991c}
W.~F\"{o}rner, Quantum and disorder effects in {D}avydov soliton theory,
  Physical Review A 44~(4) (1991) 2694--2708.
\newblock \href {https://doi.org/10.1103/PhysRevA.44.2694}
  {\path{doi:10.1103/PhysRevA.44.2694}}.

\bibitem{Pace1998}
C.~N. Pace, J.~M. Scholtz, A helix propensity scale based on experimental
  studies of peptides and proteins, Biophysical Journal 75~(1) (1998) 422--427.
\newblock \href {https://doi.org/10.1016/S0006-3495(98)77529-0}
  {\path{doi:10.1016/S0006-3495(98)77529-0}}.

\bibitem{Zakharov1972}
V.~F. Zakharov, A.~B. Shabat, Exact theory of two-dimensional self-focusing and
  one-dimensional self-modulation of wave in nonlinear media, Journal of
  Experimental and Theoretical Physics 34~(1) (1972) 62--69.

\bibitem{Kivshar1989}
Y.~S. Kivshar, B.~A. Malomed, Dynamics of solitons in nearly integrable
  systems, Reviews of Modern Physics 61~(4) (1989) 763--915.
\newblock \href {https://doi.org/10.1103/RevModPhys.61.763}
  {\path{doi:10.1103/RevModPhys.61.763}}.

\bibitem{Davydov1987}
A.~S. Davydov, V.~N. Ermakov, Linear and nonlinear resonance electron tunneling
  through a system of potential barriers, Physica D: Nonlinear Phenomena 28~(1)
  (1987) 168--180.
\newblock \href {https://doi.org/10.1016/0167-2789(87)90127-8}
  {\path{doi:10.1016/0167-2789(87)90127-8}}.

\bibitem{Ermakov1988}
V.~N. Ermakov, E.~A. Ponezha, Resonance tunneling with dissipation taken into
  account, Physica Status Solidi (b) 145~(2) (1988) 545--554.
\newblock \href {https://doi.org/10.1002/pssb.2221450220}
  {\path{doi:10.1002/pssb.2221450220}}.

\bibitem{Forinash1994}
K.~Forinash, M.~Peyrard, B.~Malomed, Interaction of discrete breathers with
  impurity modes, Physical Review E 49~(4) (1994) 3400--3411.
\newblock \href {https://doi.org/10.1103/PhysRevE.49.3400}
  {\path{doi:10.1103/PhysRevE.49.3400}}.

\bibitem{Ostrovskaya1999}
E.~A. Ostrovskaya, Y.~S. Kivshar, D.~V. Skryabin, W.~J. Firth, Stability of
  multihump optical solitons, Physical Review Letters 83~(2) (1999) 296--299.
\newblock \href {https://doi.org/10.1103/PhysRevLett.83.296}
  {\path{doi:10.1103/PhysRevLett.83.296}}.

\bibitem{Brizhik1983}
L.~S. Brizhik, A.~S. Davydov, Soliton excitations in one-dimensional molecular
  systems, Physica Status Solidi (b) 115~(2) (1983) 615--630.
\newblock \href {https://doi.org/10.1002/pssb.2221150233}
  {\path{doi:10.1002/pssb.2221150233}}.

\bibitem{Brizhik1988}
L.~S. Brizhik, Y.~B. Gaididei, A.~A. Vakhnenko, V.~A. Vakhnenko, Soliton
  generation in semi-infinite molecular chains, Physica Status Solidi (b)
  146~(2) (1988) 605--612.
\newblock \href {https://doi.org/10.1002/pssb.2221460221}
  {\path{doi:10.1002/pssb.2221460221}}.

\bibitem{Brizhik1993}
L.~S. Brizhik, Soliton generation in molecular chains, Physical Review B 48~(5)
  (1993) 3142--3144.
\newblock \href {https://doi.org/10.1103/PhysRevB.48.3142}
  {\path{doi:10.1103/PhysRevB.48.3142}}.

\bibitem{Kuprievich1990}
V.~A. Kuprievich, On the calculations of the exciton-phonon coupling parameters
  in the theory of {D}avydov solitons, in: P.~L. Christiansen, A.~C. Scott
  (Eds.), Davydov's Soliton Revisited: Self-Trapping of Vibrational Energy in
  Protein, Springer, New York, 1990, pp. 199--207.
\newblock \href {https://doi.org/10.1007/978-1-4757-9948-4_15}
  {\path{doi:10.1007/978-1-4757-9948-4_15}}.

\bibitem{Devault1984}
D.~De~Vault, Quantum-Mechanical Tunneling in Biological Systems, Vol.~2,
  Cambridge University Press, Cambridge, UK, 1984.

\bibitem{Brookes2017}
J.~C. Brookes, Quantum effects in biology: golden rule in enzymes, olfaction,
  photosynthesis and magnetodetection, Proceedings of the Royal Society A 473
  (2017) 20160822.
\newblock \href {https://doi.org/10.1098/rspa.2016.0822}
  {\path{doi:10.1098/rspa.2016.0822}}.

\bibitem{Klinman2013}
J.~P. Klinman, A.~Kohen, Hydrogen tunneling links protein dynamics to enzyme
  catalysis, Annual Review of Biochemistry 82~(1) (2013) 471--496.
\newblock \href {https://doi.org/10.1146/annurev-biochem-051710-133623}
  {\path{doi:10.1146/annurev-biochem-051710-133623}}.

\bibitem{Basran2001}
J.~Basran, S.~Patel, M.~J. Sutcliffe, N.~S. Scrutton, Importance of barrier
  shape in enzyme-catalyzed reactions: vibrationally assisted hydrogen
  tunneling in tryptophan tryptophylquinone-dependent amine dehydrogenases,
  Journal of Biological Chemistry 276~(9) (2001) 6234--6242.
\newblock \href {https://doi.org/10.1074/jbc.M008141200}
  {\path{doi:10.1074/jbc.M008141200}}.

\bibitem{Sutcliffe2000}
M.~J. Sutcliffe, N.~S. Scrutton, Enzyme catalysis: over-the-barrier or
  through-the-barrier?, Trends in Biochemical Sciences 25~(9) (2000) 405--408.
\newblock \href {https://doi.org/10.1016/S0968-0004(00)01642-X}
  {\path{doi:10.1016/S0968-0004(00)01642-X}}.

\bibitem{GeorgievGlazebrook2018}
D.~D. Georgiev, J.~F. Glazebrook, The quantum physics of synaptic communication
  via the {SNARE} protein complex, Progress in Biophysics and Molecular Biology
  135 (2018) 16--29.
\newblock \href {https://doi.org/10.1016/j.pbiomolbio.2018.01.006}
  {\path{doi:10.1016/j.pbiomolbio.2018.01.006}}.

\end{thebibliography}

\end{document}